\documentclass[journal,12pt,draftclsnofoot,onecolumn]{IEEEtran}
\usepackage{amsmath,amsfonts}
\usepackage{algorithmic}
\usepackage{algorithm}
\usepackage{array}
\usepackage[font=small,labelfont=rm,textfont=rm]{caption} % 设置全局 caption 字体大小
\usepackage[font=small,labelfont=rm,textfont=rm]{subfig}
\usepackage{textcomp}
\usepackage{stfloats}
\usepackage{url}
\usepackage{verbatim}
\usepackage{graphicx}
\usepackage{cite}
\usepackage{multirow}  %复杂表格；
\usepackage{makecell}
\usepackage{diagbox}  %引入diagbox宏包
\usepackage{color}
\graphicspath{{./Figures/}}	
\makeatletter
\renewcommand{\maketag@@@}[1]{\hbox{\m@th\normalsize\normalfont#1}}%
\makeatother %公式字体变小时将编号恢复原样

\begin{document}

\title{Performance Boundaries and Tradeoffs in Super-Resolution Imaging Technologies for Space Targets}

\author{Xiaole He, Ping Liu, Junling Wang*}
        % <-this % stops a space

%\markboth{IEEE TRANSACTIONS ON Aerospace and Electronic Systems,~Vol.~XX, No.~X, Nov.~2024}%

\markboth{XXX}%
{Shell \MakeLowercase{\textit{et al.}}: A Sample Article Using IEEEtran.cls for IEEE Journals}

\IEEEpubid{0000--0000/00\$00.00~\copyright~2021 IEEE}

\maketitle
\begin{abstract}
Inverse synthetic aperture radar (ISAR) super-resolution imaging technology is widely applied in space target imaging. However, the performance limits of super-resolution imaging algorithms remain a rarely explored issue. This paper investigates these limits by analyzing the boundaries of super-resolution algorithms for space targets and examines the relationships between key contributing factors. In particular, drawing on the established mathematical theory of computational resolution limits (CRL) for line spectrum reconstruction, we derive mathematical expressions for the upper and lower bounds of cross-range super-resolution imaging, based on ISAR imaging model transformations. 
Leveraging the explicit expressions, we first explore influencing factors of these bounds, such as the traditional Rayleigh limit, the number of scatterers, and the peak signal-to-noise ratio (PSNR) of scatterers. Then we elucidate the minimum resource requirements in ISAR imaging imposed by the CRL theory to meet the desired cross-range resolution, without which studying super-resolution algorithms becomes unnecessary in practice. Furthermore, the tradeoffs between the cumulative rotation angle, the radar transmit energy, and other contributing factors in optimizing the resolution are analyzed. Simulations are conducted to demonstrate these tradeoffs across various ISAR imaging scenarios, revealing their high dependence on specific imaging targets.  
\end{abstract}


\section{Introduction}
\IEEEPARstart{I}{nverse} Synthetic Aperture Radar (ISAR) imaging techniques reconstruct high-resolution 2-D images by exploiting the Doppler differences of target scatterers observed by wideband radar. These techniques are widely applied in both military and civilian fields, including target identification, aircraft traffic control, and air/space surveillance \cite{ref1}. 

The most common ISAR imaging technique is the Range-Doppler (RD) algorithm \cite{ref2}. This time-frequency representation (TFR) method assumes that the space target conforms to a uniform rotation during the coherent processing interval (CPI), allowing the radar observation process to be modeled as a standard Fourier transform (FT) from the echoes to the ISAR image \cite{ref3}.
However, the space target adheres to a non-uniform rotation, causing the nonlinear Doppler effects and cross terms in migration through resolution cell (MTRC) that blur the imaging results produced by the RD algorithm. To address these limitations, alternative TFR methods have been proposed. The range instantaneous Doppler (RID) method in \cite{ref4,ref5,ref6,ref7} replaces linear Doppler analysis with a high-order spectrum transform kernel that is insensitive to the target's time-variant phase terms, enabling the analysis of instantaneous Doppler shifts. The fractional and local polynomial Fourier transform methods \cite{ref8,ref9,ref10,ref11} extend the RD and RID approaches, offering more precise time-frequency analysis. The synchrosqueezing transformation method \cite{ref12,ref13} replaces FT with continuous wavelet transform (CWT) or short-time Fourier transform (STFT), producing sharper TFR by realigning the transform outcomes. Additionally, methods such as the adaptive joint time-frequency transform \cite{ref14} and the S-method \cite{ref15} are commonly used in TFR analysis. While these methods are generally simple and effective, a compromise between the CPI, TFR cross-term suppression, and cross-range resolution is unavoidable in one form or another \cite{ref16}.

Several compensation techniques have been developed to address non-uniform rotation and MTRC in space target ISAR imaging, simplifying the complex imaging problem into a linear system suitable for FT processing. These techniques can be broadly categorized into parametric and non-parametric methods. Parametric methods estimate the target's motion parameters through mathematical modeling, enabling compensation for the phase errors compensation induced by non-uniform rotation. For example, in \cite{ref17}, parameters such as relative angular acceleration and relative angular jerk are estimated to compensate for non-uniform rotation, resulting in well-focused ISAR images. In contrast, non-parametric methods directly extract motion information from the echo signal to perform phase compensation without constructing a specific motion model. The Matching Fourier Transform (MFT) technique \cite{ref18} compensates for the quadratic phase components caused by accelerated motion. The linear Keystone transform \cite{ref19} is a common tool for MTRC correction, and eliminates the coupling between fast-frequency and slow-time through linear interpolation. However, it does not account for high-order phase terms caused by non-uniform motion. To address thies challenge, various improved Keystone transform algorithms have been proposed to handle the complex motion of space targets in ISAR imaging \cite{ref20,ref21,ref22,ref23,ref24}. 
For instance, \cite{ref20} models target motion as a quartic function, introduces a fourth-order Keystone transform to eliminate coupling between high-order phase terms and slow time, and proposes an MTRC-AHP algorithm to correct MTRC and compensate for cross-range high-order phase errors.

\IEEEpubidadjcol
Super-resolution algorithms provide another effective solution for the nonlinear imaging problem and can be divided into two main categories: point-source and distributed-source techniques \cite{ref25}. 
Point-source techniques treat the ISAR image as a set of finite discrete points, aiming to reconstruct their locations and amplitudes. Typical algorithms include MUSIC, ESPRIT, CLEAN, RELAX, and Bandwidth Extrapolation(BWE). Distributed-Source techniques, on the other hand, model the echo signal from each scatterer as a continuous high-resolution function over a regular sampling grid, filtered by the Point Spread Function (PSF) of the imaging system. Typical algorithms include the Least Square (LS), Singular Value Decomposition (SVD), Capon's Minimum Variance Method (MVM), and Amplitude and Phase Estimation of a Sinusoid (APES). Furthermore, the Compressed Sense (CS) algorithm, which exploits the sparse representation of ISAR images in the Fourier domain, has proven to be effective for super-resolution imaging \cite{ref26,ref27,ref28,ref29}.
While these algorithms have been successful in reconstructing clear ISAR images across various applications, their performance limits are rarely discussed.

The Rayleigh Criterion \cite{ref30} is commonly employed by TFR methods to evaluate the resolvability of two adjacent point sources. According to this criterion, two-point sources are considered distinguishable if the minimum distance between their PSFs exceeds half the width of the main lobe. 
In \cite{ref31}, the feasibility of distinguishing adjacent point sources based on the Rayleigh Criterion was examined, and the minimum resolvable distance for stable reconstruction, known as the Rayleigh limit (RL), was determined. Additionally, \cite{ref31} investigated the possibility of stably recovering point sources from noisy data below the RL, a challenge known as the super-resolution problem.

In the analysis of resolution limits for super-resolution algorithms, some researchers emphasize the stability of specific algorithms to ensure their reliability in practical applications. In \cite{ref32}, an explicit resolution estimation of the MUSIC algorithm was provided, considering the perturbation of the noise-space correlation function when frequencies are pairwise separated by more than two Rayleigh limits in each direction. Several studies \cite{ref33,ref34,ref35} examined the stability of the MUSIC and ESPRIT algorithms in spectrum estimation problems under finite sampling and Gaussian white noise models, even when the minimum distance between adjacent point sources is below the RL.
In \cite{ref36}, a compressive blind array calibration method using ESPRIT with random sub-arrays was proposed, relaxing the signal model restrictions and providing a non-asymptotic error bound. 
Additionally, \cite{ref37} and \cite{ref38} derived the upper bound of the minimum resolvable distance required for stable recovery utilizing the LASSO algorithm and its improved variant, the BLASSO algorithm. 
Other researchers focused on the intrinsic super-resolution capabilities of the imaging problem itself. Studies \cite{ref39,ref40,ref41,ref42,ref43} assumed specific distributions of point sources and applied mathematical tools to derive the minimax error for stable recovery. However, these theoretical error bounds rely on an indeterminate positive constant and cannot be directly applied to practical imaging problems.

Current research on super-resolution algorithm performance primarily focuses on scenarios with two-point sources. However, ISAR imaging targets typically have complicated structures involving more than two scatterers within a range cell, and the performance limits of super-resolution algorithms in such cases are rarely discussed.
In studies\cite{ref46,ref48,liu2023super}, researchers introduced the concept of "computational resolution limits", which define the minimum required distance between point sources to accurately resolve their number and locations under specific noise levels. By employing nonlinear approximation theory in Vandermonde space, the theoretical computational resolution limit for signal reconstruction was successfully derived. 
The findings in \cite{ref46} are particularly notable for providing a clear boundary value without indeterminate constants, making them directly applicable to practical imaging scenarios. Unfortunately, these results are based on uniform sampling and single frequency assumptions, limiting their application to ISAR imaging of space targets, which often exhibit nonlinear behaviors such as non-uniform rotation and sampling. To address this, we employ the parameterized MTRC-AHP compensation algorithm in \cite{ref20}, transforming the space target imaging problem into a linear system. 
This allows the analysis of the resolution limit in ISAR imaging using the mathematical tools from \cite{ref46}, offering a more comprehensive understanding of the applications and limitations of ISAR imaging algorithms. Moreover, with advancements in phased array technology, radar beam resources can be allocated more efficiently based on this resolution limit, thereby enhancing the imaging efficiency of wideband radars.

Therefore, this paper investigates the theoretical boundaries of super-resolution imaging algorithms for space targets, as well as the key influencing factors and their associated tradeoffs. The primary novelties and contributions of this work are as follows:
\begin{itemize}
	\item [1)] 
    After a model transformation from the nonlinear ISAR imaging problem into a linear spectrum estimation problem, we derive mathematical expressions for the upper and lower bounds of cross-range resolution in ISAR imaging, based on the latest computational resolution limit theory of the line spectrum reconstruction problem.
	\item [2)]
    We establish an explicit relationship between the performance of ISAR super-resolution imaging algorithms, the traditional Rayleigh limit, the number of scatterers, and the peak SNR of scatterers, facilitating a comprehensive analysis of tradeoffs and constraints among these factors in ISAR imaging.
	\item [3)]
    We analyze the minimum resource requirements for the desired cross-range resolution and explore the tradeoffs among various influencing factors across different space target ISAR imaging scenarios. Through detailed simulations, we assess the impact of these tradeoffs on imaging efficiency, providing insights into optimizing resource configuration for enhanced imaging performance.
\end{itemize}
  
The remainder of this study is organized as follows. 
In Section \ref{section:signal-model}, the nonlinear ISAR imaging problem for space targets is transformed into a linear spectrum estimation problem using the MTRC-AHP algorithm \cite{ref20}. We then derive the computational resolution limits of super-resolution ISAR imaging algorithms and compare them with the Rayleigh limit.
Section \ref{section:bounds} discusses tradeoffs and performance limits of influencing factors in detail based on the derived resolution boundaries.
Section \ref{section:simulation} illustrates the relationship between the super-resolution limit and the Rayleigh limit, explores the tradeoffs among various factors, and examines their impact on imaging efficiency across different scenarios. Finally, Section \ref{section:conclusion} concludes the study. 

\section{ISAR imaging algorithm performance limit} \label{section:signal-model}
\subsection{Space Target ISAR Imaging to Stable Recovery Problem}

Assume a wideband radar is employed to observe a space target, emitting a linear frequency modulation (LFM) signal with a bandwidth of $B_w$. After translation motion compensation and range compression, the one-dimensional (1-D) range profile of scatterer $q$ at slow time $t_n$ can be expressed as:
\begin{equation}
	\begin{aligned} 
			s\left( \hat{t},{{t}_{n}} \right)={{A}_{q}}\text{sinc}\left[ {{B}_{w}}\left( \hat{t}-\frac{{{R}_{q}}({{t}_{n}})}{c} \right) \right]\exp \left[ -j2\text{ }\!\!\pi\!\!\text{ }{{f}_{c}}\frac{{{R}_{q}}({{t}_{n}})}{c} \right]
			\label{equ:1Drangeprofile}
		\end{aligned}
\end{equation}

\noindent where $\operatorname{sinc}\left( x \right)={\sin \left( \pi x \right)}/{\pi x}$, $A_{q}$ is the scattering coefficient of scatterer $q$ after range compression, $f_{c}$ and $c$ respectively represent the carrier frequency and the speed of light, ${{R}_{q}}\left( {{t}_{n}} \right)$ is the instantaneous round-trip range of scatterer q relative to the target center along the radar line of sight (LOS) at $t_n$.

It should be noted here that we assume the electromagnetic scattering phases of each scatterer are identical in (\ref{equ:1Drangeprofile}). This assumption is made because phase differences among scatterers, excluding those induced by scatterer positions, lead to a better theoretical super-resolution limit, as illustrated in the two-point case discussed in \cite{ref48}. 
In fact, most space targets are predominantly composed of components with repetitive structures and similar geometric layouts, such as satellite solar panels, antenna arrays, and bolts, the scattering characteristic-induced phases of scatterers on these components are generally identical due to their shared viewing angle in a specific ISAR image. Consequently, assuming identical phases is more realistic when considering the existence of these components to derive universal super-resolution limits in space target ISAR imaging.

\begin{figure}[!h]
	\centering
	\includegraphics[width=2.5in]{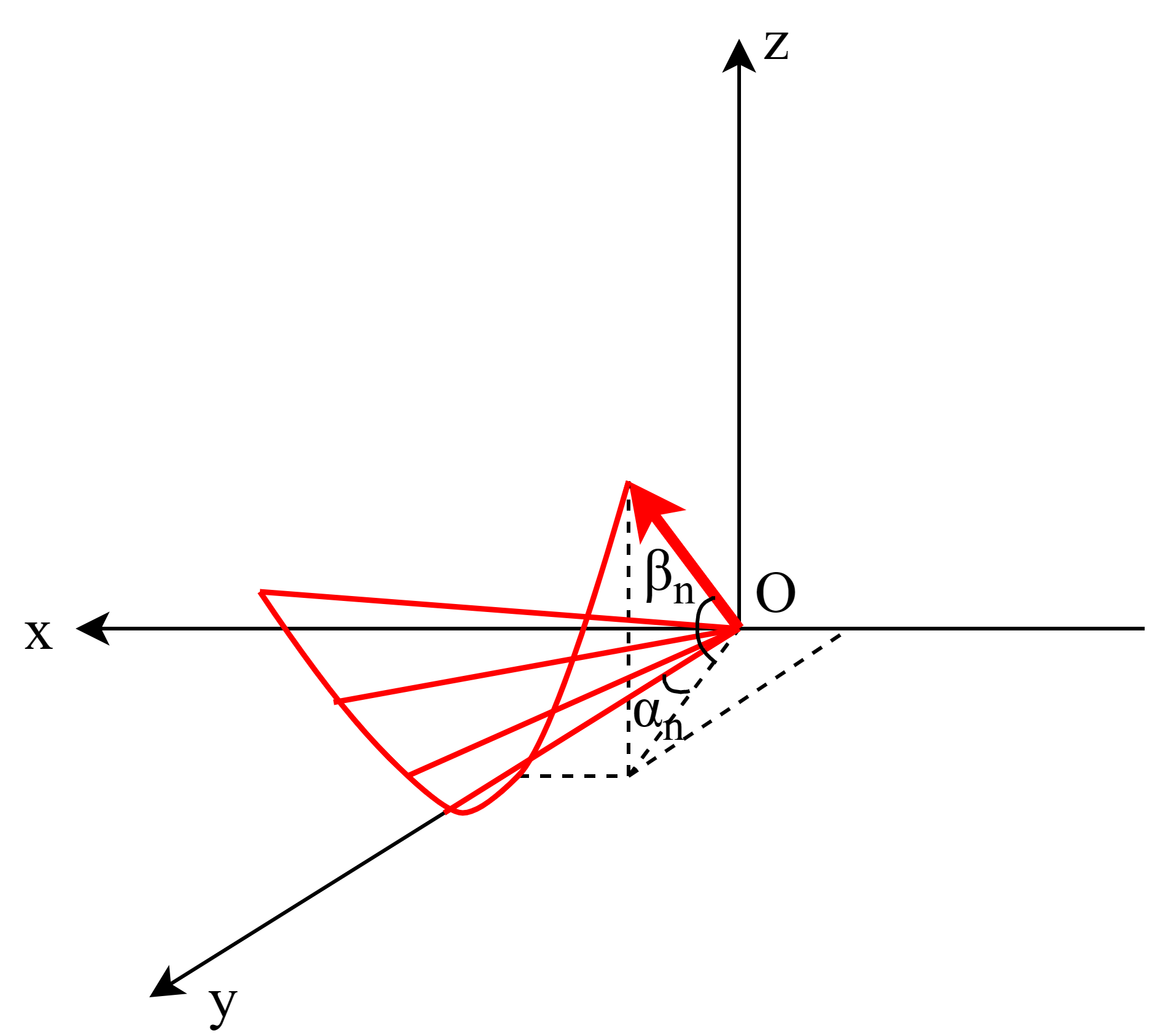}
	\caption{Schematic diagram of the change of radar LOS in $T_{\mathrm{Img0}}$}
	\label{fig:LOSvariation}
\end{figure}

Fig.\ref{fig:LOSvariation} shows the trajectory of radar LOS in the reference imaging coordinate system $T_{\mathrm{Img0}}$ during the CPI of one ISAR image \cite{ref20}. The $T_{\mathrm{Img0}}$ system is defined as the imaging coordinate system $T_{\mathrm{Img}}\left(x_{\text{im}},y_{\text{im}},z_{\text{im}}\right)$ at the intermediate time of CPI, $t_0$. In the $T_{\mathrm{Img}}\left(x_{\text{im}},y_{\text{im}},z_{\text{im}}\right)$ system, the origin $O$ is the target center, the $y_{\text{im}}$-axis aligns with the radar LOS, known as the range direction of the ISAR image, the $z_{\text{im}}$-axis points towards the direction of the target’s effective rotation vector, and the $x_{\text{im}}$-axis, known as the cross-range or azimuth direction of the ISAR image, forms a right-hand Cartesian coordinate system with the $y_{\text{im}}$-axis and $z_{\text{im}}$-axis. Since the radar LOS rotates around the target center on a highly nonlinear curved surface, the ISAR imaging plane, formed by the $x_{\text{im}}$-axis and $y_{\text{im}}$-axis, exhibits spatial variation. 

In $T_{\mathrm{Img0}}\left(x,y,z\right)$, assuming $r_{q}=[x_{q},y_{q},z_{q}]$ represent the coordinates of scatterer $q$ on the target. Consequently, ${{R}_{q}}\left( {{t}_{n}} \right)$  can be expressed as:
\begin{equation}
	\begin{aligned}  %让公式和编号在同一行，仅缩小公式
			{{R}_{q}}\left( {{t}_{n}} \right)=2\left( {{x}_{q}}\cos {{\beta }_{n}}\sin {{\alpha }_{n}}+{{y}_{q}}\cos {{\beta }_{n}}\cos {{\alpha }_{n}}+{{z}_{q}}\sin {{\beta }_{n}} \right)
		\end{aligned}
	\label{equ:roundtripfunction}
\end{equation}

\noindent where $\alpha_n$ and $\beta_n$ represent the azimuth and elevation angles of radar LOS at $t_n$ depicted in Fig.\ref{fig:LOSvariation}, respectively. Identity (\ref{equ:roundtripfunction}) decomposes the ${{R}_{q}}\left( {{t}_{n}} \right)$ into the sum of coordinates of scatterer $q$, each weighted by trigonometric functions of $\alpha_n$ and $\beta_n$.

Although the pulse signals are transmitted at a fixed pulse repetition frequency (PRF),  $R_q\left( t_n \right)$ exhibits nonlinearity concerning $t_n$ in both the azimuth and elevation dimensions. This nonlinearity arises primarily from two factors: the non-uniform translational motion between the target and radar, and the spatial variation of the ISAR imaging plane. Additionally, as indicated in (\ref{equ:roundtripfunction}), imaging scenarios with larger target size or larger cumulative angles will aggravate this nonlinearity. As a result, the highly nonlinear property of ${{R}_{q}}\left( {{t}_{n}} \right)$ complicates the application of classical spectrum analysis or signal reconstruction techniques, making it difficult to accurately estimate the location, amplitude, and phase information of the target scatterers.

After translation motion compensation, the MTRC-AHP algorithm proposed in \cite{ref20} can be applied to mitigate this nonlinearity. In the resulting 1-D range profiles obtained using the MTRC-AHP algorithm, ${{R}_{q}}\left( {{t}_{n}} \right)$ becomes uniform in virtual slow time $\tau_n$ and the phase becomes linearly related to the scatterer coordinates. The new 1-D range profile of scatterer $q$ can be approximated as:
\begin{equation}
	\begin{aligned}
		s\left( \hat{t},{{\tau }_{n}} \right) = {{A}_{q}}  \cdot \text{sinc}\left[ {{B}_{w }}\left( \hat{t}-\frac{2{{y}_{q}}}{c} \right) \right]  
		 \cdot \exp \left( -j\frac{4\pi }{\lambda }{{x}_{q}}{{\omega}_{a}}{{\tau }_{n}} \right)  
		 \cdot \exp \left( -j\frac{4\pi }{\lambda }{{y}_{q}} \right)
	\end{aligned}
	\label{equ:new1Drangeprofile}
\end{equation}

\noindent where $\lambda$ is the carrier wavelength, $\omega_{a}$ is the angular velocity of $\alpha_n$,  $\tau_n$ stands for the uniform virtual slow time, defined as $\tau_n=-T_{\text{CPI}}/2+n\cdot T_{\text{PRT}},n=[0,1,\cdots, N]$, where ${{T}_{\text{CPI}}}=N\cdot {{T}_{\text{PRT}}}$ is the CPI, ${{T}_{\text{PRT}}} = 1/F_{\text{PRF}}$, $F_{\text{PRF}}$ is the PRF.

Assuming there are $K$ scatterers within the $m$th range cell, the scatterer set within this range cell can be modeled as:
\begin{equation}
	\mu =\sum\limits_{q=1}^{K}{{{a}_{q}}{{\delta }_{{{{\tilde{x}}}_{q}}}}},{{a}_{q}}>0
	\label{equ:scatterersset}
\end{equation}
where ${{\tilde{x}}_{q}}$ is the equivalent azimuth coordinate of scatterer $q$, defined as:
\begin{equation}
	\tilde{x}_q=\frac{2{\omega}_a}{\lambda}x_q
	\label{equ:equivalentx}
\end{equation}

According to (\ref{equ:new1Drangeprofile}), the noisy signal echo of scatterer $q$ in the $m$th range cell can be rewritten as follows:
\begin{align}
   y\left( {{\tau }_{n}} \right)=\sum\limits_{q=1}^{K}{{{{\tilde{A}}}_{q}}{{e}^{-j2\pi {{{\tilde{x}}}_{q}}{{\tau }_{n}}}}}+w \left( {{\tau }_{n}} \right),\quad {{\tau }_{n}}\in \left[ -\frac{T_{\text{CPI}}}{2}\;,\frac{T_{\text{CPI}}}{2}\; \right]
   \label{equ:simplfiedecho}
\end{align}
where ${{\tilde{A}}_{q}}={{A}_{q}}\text{sinc}\left[ {B_w}\left( {{{\hat{t}}}_{m}}-\frac{2{{y}_{q}}}{c} \right) \right] \exp\left( -j\frac{4\pi }{\lambda }{{y}_{q}} \right)$, $w \left( {{\tau }_{n}} \right)$ is the noise, which satisfies $\left| w \left( {{\tau }_{n}} \right) \right|<\sigma$. 
 
Performing FT on  $y\left( {{\tau }_{n}}\right)$ along with $\tau_{n}$ yields the cross-range profile in $m$th range cell \cite{ref20}:

\begin{equation}
 	Y\left( {{f}_{d}} \right)=\sum\limits_{q=1}^{K}{{{a}_{q}}\operatorname{sinc}\left[ {{T}_{\text{CPI}}}\left( {{f}_{d}}-{{{\tilde{x}}}_{q}} \right) \right]}+W\left( {{f}_{d}} \right)
 	\label{equ:crossrangeprofile}
\end{equation}
where $a_q$ represents the scattering coefficient of scatterer $q$ in Doppler domain, $f_{d}$ is the Doppler frequency with a distribution range of $[-F_{\text{PRF}}/2, F_{\text{PRF}}/2]$, $W(f_d)$ is the Doppler dimension noise. Additionally, we define: 
 \begin{equation}
 	{{a}_{\min }}=\underset{q=1,\cdots ,K}{\mathop{\min }}\,\left| {{a}_{q}} \right|
 	\label{equ:minimumamp}
 \end{equation}
 
According to (\ref{equ:new1Drangeprofile}), (\ref{equ:simplfiedecho}), and (\ref{equ:crossrangeprofile}), after MTRC-AHP compensation, the location information $\{\tilde{x}\}$ of each scatterer is mapped to the Doppler frequency $\{f_d\}$, corresponding to the peak of the $\text{sinc}$ function envelope in the Doppler spectrum.  Hence, the ISAR cross-range imaging problem can be restated as a spectral estimation problem, where the goal is to estimate the scatterer locations $\{\tilde{x}\}$ from the uniformly sampled signal $y(\tau_n)$.

\subsection{Resolution Limit of ISAR Cross-Range Imaging}
The primary objective in analyzing the resolvability of adjacent scatterers using spectrum estimation techniques is to determine the minimum separation required for stable recovery of each scatterer, denoted by $D$ and defined as:
\begin{equation}
	D=\underset{p\ne q}{\mathop{\min }}\,\left| {{{\tilde{x}}}_{p}}-{{{\tilde{x}}}_{q}} \right|
	\label{equ:D}
\end{equation}

Substituting  (\ref{equ:equivalentx}) into (\ref{equ:D}) yields the relationship between the cross-range resolution of the ISAR image and $D$:
\begin{equation}
	\delta a=\frac{{\lambda}}{2{{\omega}_{a}}}D
	\label{equ:crossrangeresolution}
\end{equation}
where $\delta{a}=\underset{p\ne q}{\mathop{\min }}\,\left| {{x}_{p}}-{{x}_{q}} \right|$ represents the cross-range resolution of 2-D ISAR image.

Referring to (\ref{equ:D}) and (\ref{equ:crossrangeresolution}), the equivalent minimum separation $D$ between scatterers is proportional to the cross-range resolution $\delta{a}$ of 2-D ISAR image. For this reason, by analyzing the theoretical resolution limit of the spectral estimation problem (\ref{equ:simplfiedecho}), the azimuth resolution limit of the ISAR imaging problem (\ref{equ:new1Drangeprofile}) is obtained.

\vspace{0.2cm} 
\noindent\textbf{Rayleigh Limit}
\vspace{0.2cm} 

The resolution limit of the spectral estimation problem is usually measured by the Rayleigh criterion. 
\cite{ref31} discussed the problem of recovering a measure $\mu$ from its noisy Fourier data $y\left(w\right)$ (with cutoff frequency $\Omega \ge \left| w \right|$) in detail, and stated the modified Rayleigh criterion: point like sources separated by at least $\Delta$ can be resolved into separate sources, provided ${\pi }/{\Delta }\leq \Omega $. Therefore, following the same way as \cite{ref31}, we regard $\frac{\pi}{\Omega}$ as the RL.
Note that the common model utilized in \cite{ref31} aligns with the imaging problem presented in (\ref{equ:simplfiedecho}), indicating that the $\omega$ domain in \cite{ref31} corresponds to the $\tau_n$ domain in this work. Consequently, the cutoff frequency $\Omega$ of mathematical model (\ref{equ:simplfiedecho}) is defined in the $\tau_n$ domain as $\Omega= 2\pi (\tau_{n})_{\text{max}}  = \pi T_{\text{CPI}}$. Furthermore, the scatterer set $\mu$ is resolvable in its dual domain, the $f_d$ domain. As illustrated in (\ref{equ:crossrangeprofile}), the band-limited PSF of the spectral line is given by:
\begin{equation}
f\left(\tilde{x}_q\right)=\text{sinc}\left[T_{\text{CPI}}\left(f_d-\tilde{x}_q \right) \right] 
	\label{equ:PSF}
\end{equation}

Accordingly, the Rayleigh limit for the stable recovery of scatterer set $\mu$ from its noisy data $y(\tau_n)$, denoted by $D_{\text{RL}}$, is given by:
\begin{equation}
	D_{\text{RL}} = \frac{\pi}{\Omega} = \frac{1}{T_{\text{CPI}}} 
	\label{equ:DRL}
\end{equation}

Substituting  (\ref{equ:DRL}) into (\ref{equ:crossrangeresolution}), the Rayleigh limit for the cross-range ISAR imaging problem described in (\ref{equ:simplfiedecho}), denoted by $\delta{a}_{\text{RL}}$, can be obtained as follows, which matches the result in \cite{ref52}:
\begin{equation}
	\delta a_{\text{RL}} = \frac{\lambda}{2 \omega_a} D_{\text{RL}} = \frac{\lambda}{2\theta_{\Delta}} 
	\label{equ:crossrangeresolutionRL}
\end{equation}
where $\theta_{\Delta} = \omega_a \cdot T_{\text{CPI}}$ is the azimuth cumulative rotation angle of radar LOS during the $T_{\text{CPI}}$.

\vspace{0.2cm} 
\noindent\textbf{Computational Resolution Limit}
\vspace{0.2cm} 

However, Rayleigh's choice of resolution limit is based on the presumed resolving ability of the human visual system, which at first glance seems arbitrary \cite{ref46}. 
Recent studies \cite{li2020super,li2022stability,jordan2023fundamental} have shown that super-resolution algorithms are capable of resolving adjacent points separated by distances below the RL, suggesting that the theoretical resolvable limit for simple-frequency signals is indeed lower than the RL. 
Furthermore, studies in \cite{ref46,ref48,liu2023super} have introduced the concept of "computational resolution limits" to characterize the exact minimum required distance between multiple point sources so that their locations can be stably resolved under certain noise level.

Specifically, to ensure accurate reconstruction of the scatterer set $\mu$ from its noisy Fourier data $y(\tau_n)$,  the computational resolution limit of the equivalent minimum separation between $K$ scatterers, denoted by $D_{\text{cmp}}$, must satisfy the following constraints \cite{ref46}:

\begin{equation}
	\begin{aligned}
			\frac{2e^{-1}}{\Omega} \left( \frac{\sigma}{a_{\min}} \right)^{\frac{1}{2K-1}} < D_{\text{cmp}} < \frac{2.36e\pi}{\Omega} \left( \frac{\sigma}{a_{\min}} \right)^{\frac{1}{2K-1}}
		\end{aligned}
	\label{equ:Dcmp}
\end{equation}
When $K = 2$, the upper bound of $D_{\text{cmp}}$ can be improved to:

\begin{equation}
	D_{\text{cmp}} \geq \frac{3}{\Omega} \arcsin \left( 2 \left( \frac{\sigma}{a_{\min}} \right)^{\frac{1}{3}} \right)
	\label{equ:Dcmptwopoints}
\end{equation}
The author provided a detailed explanation of this constraint in \cite{liu2023super}: stable location recovery from  $y\left( {{\tau }_{n}} \right)$ is achievable when $D_{\text{cmp}}$ exceeds the upper bound $\frac{2.36e\pi }{\Omega }{{\left( \frac{\sigma }{{{a}_{\min }}} \right)}^{\frac{1}{2K-1}}}$, and  impossible when $D_{\text{cmp}}$ falls below the lower bound $\frac{2{{e}^{-1}}}{\Omega }{{\left( \frac{\sigma }{{{a}_{\min }}} \right)}^{\frac{1}{2K-1}}}$, without any additional prior information.

Putting (\ref{equ:crossrangeresolution}) into (\ref{equ:Dcmp}),  the theoretical computational resolution limit for cross-range ISAR imaging problem described in (\ref{equ:simplfiedecho}), denoted by $\delta{a}_{\text{cmp}}$, can be constructed as:

\begin{equation}
	\begin{aligned}
			\frac{{\lambda}}{2\theta_{\Delta} }\cdot\frac{2{{e}^{-1}}}{\pi }{{\left(\frac{1}{\rho_{\text{PSNR}}} \right)}^{\frac{1}{4K-2}}}<\delta a_{\text{cmp}}<\frac{{\lambda }}{2\theta_{\Delta} }2.36e{{\left( \frac{1}{\rho_{\text{PSNR}}} \right)}^{\frac{1}{4K-2}}}
		\end{aligned}
	\label{equ:crossrangeresolutioncmp}
\end{equation}
where $\rho_{\text{PSNR}}=a^2_{\min}/\sigma^2$ is the peak signal-to-noise ratio (PSNR) of signal echo after coherent integration. 
Based on the results in \cite{liu2023super} and the previously mentioned conclusions, the lower bound of $\delta a_{\text{cmp}}$ represents the super-resolution limit that no super-resolution algorithm can surpass without additional prior information. 
In contrast, the upper bound of $\delta a_{\text{cmp}}$ defines the resolution limit that super-resolution algorithms can reliably achieve for stable scatterer recovery.

Therefore, the lower bound of $\delta a_{\text{cmp}}$ stands for the optimal resolution capability of the imaging super-resolution algorithm, known as the super-resolution limit and denoted by $\delta a_{\text{SRL}}$, i.e.,

\begin{equation}
	\delta a_{\text{SRL}} = \frac{{\lambda}}{2\theta_{\Delta} }\cdot\frac{2{{e}^{-1}}}{\pi }{{\left(\frac{1}{\rho_{\text{PSNR}}} \right)}^{\frac{1}{4K-2}}}
	\label{equ:crossrangeresolutionSRL}
\end{equation}
The upper bound of $\delta a_{\text{cmp}}$, denoted by $\delta{a}_{\text{SRU}}$, is given by:

\begin{equation}
	\delta{a}_{\text{SRU}} =
	\begin{cases}
		\dfrac{\lambda}{2\theta_{\Delta}} \dfrac{3}{\pi} \arcsin \left( 2 \left( \dfrac{1}{\rho_{\text{PSNR}}} \right)^{\frac{1}{6}} \right), & K = 2 \\[8pt]
		\dfrac{\lambda}{2\theta_{\Delta}} 2.36 e \left( \dfrac{1}{\rho_{\text{PSNR}}} \right)^{\frac{1}{4K-2}}, & K > 2
	\end{cases}
	\label{equ:crossrangeresolutionSRU}
\end{equation}
During the theoretical derivations in \cite{ref46,ref48,liu2023super} and in equations (\ref{equ:crossrangeresolutionSRL})-(\ref{equ:crossrangeresolutionSRU}), no additional assumptions regarding the super-resolution algorithm, noise statistical characteristic, or sampling interval were imposed, indicating that the MTRC-AHP interpolation operation \cite{ref20} does not affect the validity of these equations. Consequently, the lower and upper boundaries of the cross-range resolution limit apply to any super-resolution algorithm or sparse sampling imaging scenarios.
Furthermore, since the theory of computational resolution limit in \cite{ref46,ref48,liu2023super} can be generalized to imagings with other band-limited PSFs, $\delta{a}_{\text{SRL}}$ and $\delta{a}_{\text{SRU}}$ are also relevant for radar waveform design and radar speed measurement. 

In contrast to (\ref{equ:crossrangeresolutionRL}), $\delta{a}_{\text{SRU}}$ and $\delta{a}_{\text{SRL}}$ are not only dependent on $\lambda$ and $\theta_{\Delta}$, but also establish a complex nonlinear relationship with both the PSNR and the number of scatterers.
Although \cite{li2021stable, batenkov2020conditioning, batenkov2021super} also considered the effect of PSNR on the estimation of scatterers, they focus on the minimax error rate of the reconstructions and their boundaries are not as tight as those presented in \cite{ref46,ref48,liu2023super}. 
This work transforms the highly nonlinear ISAR imaging problem into a linear spectrum estimation problem using the compensation method proposed in \cite{ref20}. By applying the latest computational resolution theory, this work explicitly establishes the mathematical relationship between the performance of super-resolution ISAR imaging algorithms, the traditional Rayleigh limit, the number of scatterers, and the PSNR in ISAR imaging.

\vspace{0.1cm}
\noindent\textbf{Relationship between $\delta {{a}_{\text{RL}}}$ and $\delta {{a}_{\text{cmp}}}$ }
\vspace{0.1cm}

This section discusses the relationship between the $\delta {{a}_{\text{RL}}}$ and $\delta {{a}_{\text{cmp}}}$ in the ISAR cross-range imaging problem.

Based on the previous definitions, the ratio of the bounds for $\delta {{a}_{\text{cmp}}}$ to $\delta {{a}_{\text{RL}}}$  (excluding the specific case where $K=2$ for the upper bound) can be formulated as follows:

\begin{align}
		& {{r}_{l}}=\delta{a}_{\text{SRL}}/{\delta {{a}_{\text{RL}}}}=\frac{2{{e}^{-1}}}{\pi }{{\left( \frac{1}{{{\rho }_{\text{PSNR}}}} \right)}^{\frac{1}{4K-2}}} \notag \\ 
		& {{r}_{u}}=\delta{a}_{\text{SRU}}/{\delta {{a}_{\text{RL}}}}\;=2.36e{{\left( \frac{1}{{{\rho }_{\text{PSNR}}}} \right)}^{\frac{1}{4K-2}}} 	
	\label{equ:ratioofRLandcmp}
\end{align}

\begin{figure}[!t]
	\centering
	\subfloat[]{\includegraphics[width=2.5in]{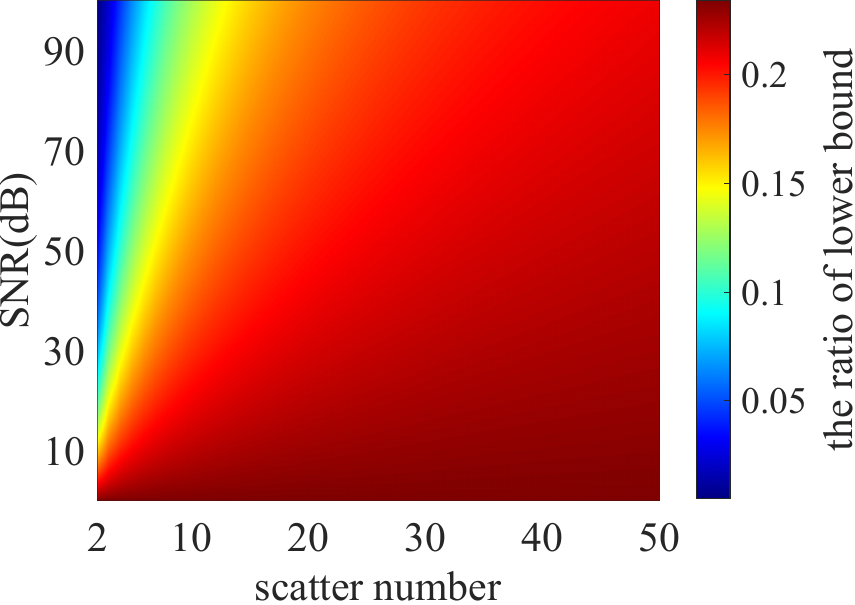}%
		\label{fig: 2a}}
	\qquad %换行
	\subfloat[]{\includegraphics[width=2.5in]{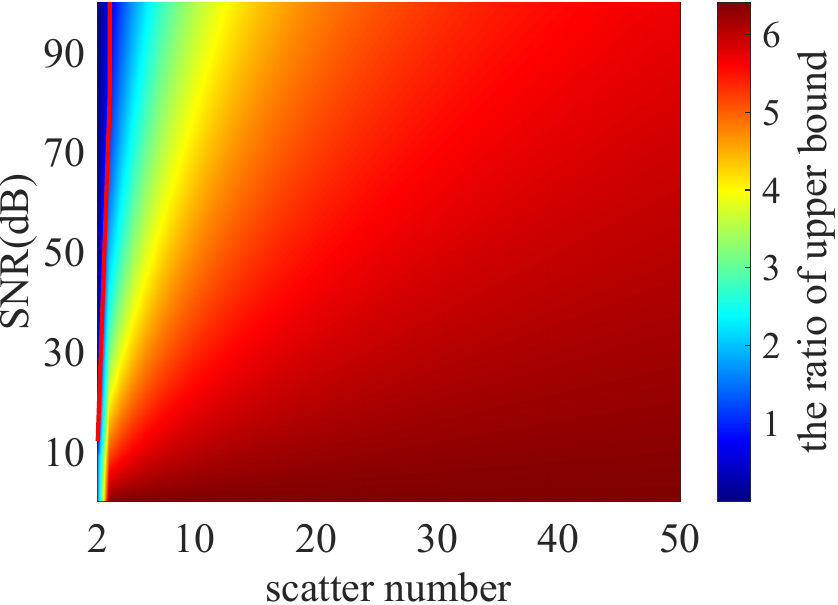}%
		\label{fig: 2b}}
	\caption{the ratio of the bounds for $\delta {{a}_{\text{cmp}}}$ to $\delta {{a}_{\text{RL}}}$. (a) the ratio of $\delta {{a}_{\text{SRL}}}$ to $\delta {{a}_{\text{RL}}}$. (b) the ratio of $\delta {{a}_{\text{SRU}}}$ to $\delta {{a}_{\text{RL}}}$.}
	\label{fig: 2}
\end{figure}

Without loss of generality, we can assume the PSNR of radar echo is greater than 0dB (i.e.,  $\rho_{\text{PSNR}} > 1$ ) after the pulse compression and FT processing in the cross-range dimension using the TFR method. 
Moreover, in typical azimuth super-resolution imaging scenarios involving space targets, there are always at least two distinct scatterers (i.e., $K\geq 2$ ). 
Let $K$ range from 2 to 50 and $\rho_{\text{PSNR}}$ range from 0dB to 100dB, the variations of $r_l$ and $r_u$ are illustrated in Fig.\ref{fig: 2}. 

As shown in Fig.\ref{fig: 2a}, $r_l$ remains consistently below 1, with the super-resolution limit ranging from approximately 0.04 to 0.22 times the Rayleigh limit. 
This finding also aligns with the results presented in \cite{ref53}, where the author introduced a novel radar waveform based on a class of self-referential interference functions, achieving a resolution up to 100 times finer than the Rayleigh limit under extremely high PSNR conditions.
Based on the simulation results in Fig.\ref{fig: 2a}, unless additional information is available, the required $\rho_{\text{PSNR}}$ to achieve the resolution reported in \cite{ref53} should be at least 83dB in practical applications. 
Fig.\ref{fig: 2b} illustrates that $r_u$ generally remains above 1. However, when $K$ is small and $\rho_{\text{SNR}}$ is high (within the red curve), there are regions where $r_u$ drops below 1.  This suggests that under these specific imaging conditions,  the optimal super-resolution algorithm can indeed achieve a more stable recovery performance compared to the Rayleigh limit.

As the focus of this study is on space target ISAR imaging, $\rho_{\text{PSNR}}$ typically falls below 50dB and the number of scatterers in a single range cell usually exceeds two. Therefore, in the ISAR imaging problem discussed in this work, it is reasonable to conclude that the Rayleigh limit of cross-range resolution satisfies the following constraints:
\begin{equation}	
	\delta {{a}_{\text{SRL}}} <\delta {{a}_{\text{RL}}} <\delta {{a}_{\text{SRU}}}
	\label{equ:inequalityofSRLSRSRU}
\end{equation}

\vspace{0.1cm}
\noindent \textbf{Guidelines for Enhancing Imaging Effiency }
\vspace{0.1cm}

According to (\ref{equ:crossrangeresolutionSRL}), the theoretical optimal cross-range resolution, $\delta {a}_{\text{SRL}}$, is determined jointly by the previously mentioned $\theta_{\Delta}$, ${\rho }_{\text{PSNR}}$, and $K$. 
Since the number of scatterers $K$  depends on the scattering characteristic of space targets, which is an intrinsic property of the target, $\theta_{\Delta}$ and ${\rho }_{\text{PSNR}}$ are intriguing factors for enhancing the theoretical resolution capability due to their tunability. 
These factors are mainly influenced by the allocated imaging resources, specifically $T_{\text{CPI}}$ and the average transmitted power $P_{\text{av}}$. 
Moreover, owing to the $\delta {a}_{\text{SRL}}$ represents the cross-range resolution limit theoretically achievable by the optimal super-resolution imaging algorithm, there is an opportunity to explore improved imaging algorithms that can allow their actual cross-range resolution, denoted by $\delta{a}_{\text{act}}$, to move towards the $\delta {a}_{\text{SRL}}$.
This section discusses the feasibility of improving the imaging algorithm as well as the necessity of enhancing the imaging resources.

In practical imaging scenarios, the desired cross-range resolution, denoted by $\delta{a}_{\text{des}}$, generally aligns with the range resolution. 
Thus, for a given configuration of imaging resources, if the theoretical optimal cross-range resolution meets the desired level while the actual cross-range resolution fails, expressed as:
\begin{align*}
    \delta {a}_{\text{SRL}} \ll \delta{a}_{\text{des}} \ll \delta{a}_{\text{act}}
\end{align*}
it is reasonable to explore a more advanced imaging algorithm to improve the actual cross-range resolution or to conserve imaging resources.

Conversely, under one imaging configuration, if the theoretical optimal cross-range resolution cannot meet the desired level, expressed as:
\begin{align*}
    \delta {a}_{\text{SRL}} \gg \delta{a}_{\text{des}}
\end{align*}
improving the super-resolution imaging algorithm becomes impractical.  In this case, increasing the imaging resources, such as extending $T_{\text{CPI}}$ or boosting $P_{\text{av}}$, is the only feasible approach to fulfill the resolution requirement. If even the maximum allocation of imaging resources provided by the radar system fails to meet the resolution requirement, the only viable strategy is to accept the resulting suboptimal cross-range resolution.

Specifically, since enhancing the imaging resources to improve resolution capability is a dynamic and interdependent process involving multiple adjustable parameters, certain resource tradeoffs exist among them. These tradeoffs will be explored thoroughly in Section \ref{subsection:tradeoff}. Notably, the available imaging resources are constrained by the radar system and ISAR imaging scenarios, which will be discussed in Section \ref{subsection:parameterlimittheory}.

\section{Influencing Factors and Performance Limits}
\label{section:bounds}

As shown in (\ref{equ:crossrangeresolutioncmp}), both the lower and upper bounds of $\delta a_{\text{cmp}}$ contain the terms $\frac{{\lambda}}{2\theta_{\Delta} }$ and ${{\left( \frac{1}{\rho_{\text{PSNR}}} \right)}^{\frac{1}{4K-2}}}$.
The first term represents the conventional Rayleigh limit, and the second term stands for a function of $\rho_{\text{PSNR}}$ and $K$, distinguishing the computational resolution limit from the Rayleigh limit.
Excluding the case of $K=2$, $\delta a_{\text{SRL}}$ and $\delta a_{\text{SRU}}$ maintain a fixed proportional relationship, indicating that the influencing factors $\theta_{\Delta}$, $\rho_{\text{PSNR}}$, and $K$,  have the same impact on both $\delta a_{\text{SRU}}$ and $\delta a_{\text{SRL}}$. 
Since $\delta a_{\text{SRL}}$ stands for the optimal achievable resolution by the most advanced super-resolution imaging algorithm, it defines the boundary between the unattainable performance region and the feasible region for radar systems. 
Therefore, analyzing $\delta a_{\text{SRL}}$ is more crucial for radar system and imaging scenario design. This section focuses on the effects of influencing factors on $\delta a_{\text{SRL}}$ and their tradeoffs. The performance limits in space target ISAR imaging are consequently provided. 

\subsection{Super-Resolution Influencing Factors} \label{section:influencinfactors}

Referring to (\ref{equ:crossrangeresolutionSRL}), holding $\rho_{\text{PSNR}}$ and $K$ constant, then the relationship between $\delta a_{\text{SRL}}$ and $\theta_{\Delta}$ can be formulated as:

\begin{equation}	
	\delta {{a}_{\text{SRL}}}\left( \theta_{\Delta}  \right)={{c}_{1}}\cdot \frac{1}{\theta_{\Delta}}
	\label{equ:crossrangeresolutionSRLthetaDelta}
\end{equation}
\begin{equation}	
	\frac{\partial \delta {{a}_{\text{SRL}}}}{\partial \theta_{\Delta} }=-c_1\frac{1}{\theta_{\Delta}^{2}}
	\label{equ:crossrangeresolutionSRLthetaDeltaderivative}
\end{equation}
where ${{c}_{1}}=\frac{{{\lambda }}}{2}\cdot \frac{2{{e}^{-1}}}{\pi }{{\left( \frac{1}{{{\rho }_{\text{PSNR}}}} \right)}^{\frac{1}{4K-2}}}$ is a positive value independent of $\theta_{\Delta}$.

As shown in (\ref{equ:crossrangeresolutionSRLthetaDelta}), $\delta {{a}_{\text{SRL}}}\left( \theta_{\Delta}  \right)$ is inversely proportional to $\theta_{\Delta} $. Therefore, an increase in $\theta_{\Delta} $ leads to a decrease in $\delta a_{\text{SRL}}$, simultaneously enhancing the resolution capability of the optimal imaging algorithm.  
As shown in (\ref{equ:crossrangeresolutionSRLthetaDeltaderivative}), the improvement rate of  $\delta {{a}_{\text{SRL}}}\left( \theta_{\Delta} \right)$ with increasing $\Delta \theta $ is inversely proportional to the square of $\theta_{\Delta} $. As a result, as $\theta_{\Delta} $ grows, the effect of the same incremental change in $\theta_{\Delta} $ on improving super-resolution capability will gradually diminish.

Similarly, based on (\ref{equ:crossrangeresolutionSRL}), with $\theta_{\Delta}$ and $K$ held constant, then the relationship between $\delta a_{\text{SRL}}$ and $\rho_{\text{PSNR}}$ can be formulated as:

\begin{equation}	
	\delta {{a}_{\text{SRL}}}\left( {{\rho }_{\text{PSNR}}} \right)={{c}_{2}}\cdot {{\left( \frac{1}{{{\rho }_{\text{PSNR}}}} \right)}^{{{c}_{3}}}}
	\label{equ:crossrangeresolutionSRLthetaPSNR}
\end{equation}
\begin{equation}	
	\frac{\partial \delta {{a}_{\text{SRL}}}}{\partial {{\rho }_{\text{PSNR}}}}=-{{c}_{2}}{{c}_{3}}{{\left( \frac{1}{{{\rho }_{\text{PSNR}}}} \right)}^{{{c}_{3}}+1}}
	\label{equ:crossrangeresolutionSRLthetaPSNRderivative}
\end{equation}
where ${{c}_{2}}=\frac{{{\lambda }}}{2\theta_{\Delta} }\cdot \frac{2{{e}^{-1}}}{\pi }$, ${{c}_{3}}=\frac{1}{4K-2}$ are positive values independent of $\rho_{\text{PSNR}}$.

As shown in (\ref{equ:crossrangeresolutionSRLthetaPSNR}), $\delta {{a}_{\text{SRL}}}\left( {\rho }_{\text{PSNR}} \right)$ is inversely proportional to the $c_3$-th power of $\rho_{\text{PSNR}}$. Therefore, an increase in $\rho_{\text{PSNR}}$ leads to a decrease in $\delta a_{\text{SRL}}$, thereby improving the resolution capability of the optimal imaging algorithm.  
As shown in (\ref{equ:crossrangeresolutionSRLthetaPSNRderivative}), the rate of decrease in $\delta {{a}_{\text{SRL}}}\left( {\rho }_{\text{PSNR}} \right)$ with increasing $\rho_{\text{PSNR}} $ is inversely proportional to the $({{c}_{3}}+1)$-th power of $\rho_{\text{PSNR}}$. 
Consequently, as $\rho_{\text{PSNR}}$ grows, the effect of the same incremental change in $\rho_{\text{PSNR}}$ on enhancing super-resolution capability gradually diminishes. 
Additionally,  increasing $K$ results in a reduction in the exponent $c_3$, in turn slowing the improvement rate of super-resolution capability. 
The polynomial increment in $\rho_{\text{PSNR}}$ corresponds to the linear improvement of $\delta a_{\text{SRL}}$, and this improvement rate is also dependent on $K$.

Following the same approach as in (\ref{equ:crossrangeresolutionSRL})  and keeping $\theta_{\Delta}$ and ${\rho }_{\text{SNR}} $ fixed, then the relationship between $\delta a_{\text{SRL}}$ and $K$ can be formulated as:

\begin{equation}	
	\delta {{a}_{\text{SRL}}}\left( K \right)={{c}_{2}}\cdot {{\left( {{c}_{4}} \right)}^{\frac{1}{4K-2}}}
	\label{equ:crossrangeresolutionSRLthetaK}
\end{equation}
\begin{equation}	
	\frac{\partial \delta {a}_{\text{SRL}}}{\partial K}=-{c_2}\cdot \ln \left( {{c}_{4}} \right)\cdot \frac{1}{{{\left( 2K-1 \right)}^{2}}}{{\left( {{c}_{4}} \right)}^{\frac{1}{4K-2}}}
	\label{equ:crossrangeresolutionSRLthetaKderivative}
\end{equation}
where ${{c}_{4}}=\frac{1}{{{\rho }_{\text{PSNR}}}}<1$ is a positive value independent of $K$.

As shown in (\ref{equ:crossrangeresolutionSRLthetaK}), $\delta {{a}_{\text{SRL}}}\left( K \right)$ increases exponentially with $K$, leading to a rapid decline in the resolution capability of the optimal imaging algorithm as $K$ increases from a small value.
However, the rate of this increase slows as $K$ grows larger, as indicated in (\ref{equ:crossrangeresolutionSRLthetaKderivative}), $\frac{\partial \delta {{a}_{\text{SRL}}}}{\partial K}$ is inversely proportional to ${{\left( 2K-1 \right)}^{2}}$. Consequently, the degradation in super-resolution capability occurs at a progressively slower rate with further increases in $K$.

\subsection{Tradeoffs between Super-Resolution Influencing Factors} \label{subsection:tradeoff} 

In the same imaging scenario, a tradeoff exists between $\theta_{\Delta}$ and ${\rho }_{\text{PSNR}}$, subject to the constraints imposed by the target's motion characteristic and the radar range equation. Understanding this tradeoff and its impact on imaging performance is crucial.

The $\theta_{\Delta}$ primarily corresponds to $T_{\text{CPI}}$, and their relationship can be formulated as:

\begin{align}
		\theta_{\Delta} =\operatorname{acos}\left( \frac{T_{\text{CPI}}^{2}{{d}_{1}}+{{T}_{\text{CPI}}}{{d}_{2}}+{{d}_{3}}}{\sqrt{T_{\text{CPI}}^{4}{{d}_{4}}+T_{\text{CPI}}^{3}{{d}_{5}}+T_{\text{CPI}}^{2}{{d}_{6}}+{{T}_{\text{CPI}}}{{d}_{7}}+{{d}_{8}}}} \right)
	\label{equ:thetaDeltawithTcon}
\end{align}

\noindent where ${{d}_{1}}\sim {{d}_{8}}$ are constants determined by the orbital elements, the radar LOS during the observation period, and the radar station's latitude and longitude.  A detailed description of these constants is provided in Appendix.\ref{appendix:appendixtheta}. It is observed that the growth of $\theta_{\Delta}$ with $T_{\text{CPI}}$ is approximately linear for each specific ISAR imaging scenario.

Assume that ${P}_{t}$ and $p_{\text{DC}}$ represent the transmitted power and the duty cycle, respectively. Then the average transmitted power, denoted by $P_{\text{av}}$, is given by $P_{\text{av}} = P_t\cdot p_{\text{DC}}$, and the transmitted energy of the radar system, denoted by $E$, is given by $E=P_{\text{av}}\cdot {{T}_{\text{CPI}}} $.
According to the radar range equation in \cite{ref54}, the ${\rho }_{\text{PSNR}}$ can be constructed as:

\begin{equation}
\rho_{\text{PSNR}} = \frac{{{C}_{\text{st}}}\cdot {{\sigma_{\text{RCS}} }}\cdot E}{{R^4}}
	\label{equ:PSNRfucntion}
\end{equation}
where $\sigma_{\text{RCS}} $ represents the radar cross section (RCS)  of the target, ${C}_{\text{st}}$ is influenced by multiple factors, with a detailed description provided in Appendix.\ref{appendix:appendixPSNR}.
Note that ${C}_{\text{st}}$ is determined by factors such as radar system parameters and propagation effects, and $\sigma_{\text{RCS}} $ depends on factors like the observation perspective, the scattering characteristics of the target, and the radar wavelength. Although both ${C}_{\text{st}}$ and $\sigma_{\text{RCS}}$ are multi-variable functions, they can be treated as constants in a specific imaging scenario.

Based on (\ref{equ:thetaDeltawithTcon}) and (\ref{equ:PSNRfucntion}), the tradeoff between $\theta_{\Delta}$ and ${\rho }_{\text{PSNR}}$ can be converted to a tradeoff between $T_{\text{CPI}}$ and ${P}_{\text{av}}$ in practical imaging scenario. This tradeoff aids in selecting the most appropriate imaging strategy for different scenarios, thereby optimizing overall imaging efficiency.

Substituting (\ref{equ:thetaDeltawithTcon}) and (\ref{equ:PSNRfucntion}) into (\ref{equ:crossrangeresolutionSRL})  yields the explicit expression of $\delta {{a}_{\text{SRL}}}$:

\begin{align}
	\delta {{a}_{\text{SRL}}} & =\frac{{{\lambda }}}{2\theta_{\Delta} }\cdot \frac{2{{e}^{-1}}}{\pi }{{ \left(\frac{R^4}{{C_{\text{st}}}\cdot {\sigma }\cdot E}\right)}^{\frac{1}{4K-2}}}\notag \\
	& = \frac{\lambda }{\operatorname{acos}\left( \frac{T_{\text{CPI}}^{2}{{d}_{1}}+{{T}_{\text{CPI}}}{{d}_{2}}+{{d}_{3}}}{\sqrt{T_{\text{CPI}}^{4}{{d}_{4}}+T_{\text{CPI}}^{3}{{d}_{5}}+T_{\text{CPI}}^{2}{{d}_{6}}+{{T}_{\text{CPI}}}{{d}_{7}}+{{d}_{8}}}} \right) } \cdot \frac{e^{-1}}{\pi}\left(\frac{R^4}{C_{\text{st}}\cdot {\sigma }\cdot E}\right)^{\frac{1}{4K-2}}
	\label{equ:SRLwithTconandE}
\end{align}
when $E$ holds constant, $p_{\text{DC}}$ and $T_{\text{CPI}}$ are constrained by:

\begin{equation}
	{{T}_{\text{CPI}}}=\frac{E}{{{P}_{t}}\cdot {{p}_{\text{DC}}}}
	\label{equ:TconwithE}
\end{equation}

As indicated in (\ref{equ:SRLwithTconandE}) and (\ref{equ:TconwithE}), for space targets in the same orbit, reducing $p_{\text{DC}}$ to extend $T_{\text{CPI}}$ is an effective approach for improving imaging performance without additional energy resources. This effectively increases  $\theta_{\Delta}$  while keeping $E$ constant. 
By comparison, when improving imaging performance by increasing $E$ while keeping $T_{\text{CPI}}$ constant, the resulting enhancement of $\delta {{a}_{\text{SRL}}}$ follows $E^{-1/(4K-2)}$. As $K$  increases, the exponent term $1/(4K-2)$ approaches zero, causing $\delta {{a}_{\text{SRL}}}$ to converge toward the RL. At this stage, increasing $E$ yields only marginal improvement in imaging resolution, making it the least favorable option.

When the super-resolution limit, the scattering characteristic of scatterers, and radar system parameters are held constant, the tradeoff between $P_t$, $p_{\text{DC}}$ and $T_{\text{CPI}}$ in a specific imaging scenario can be expressed as:

 %small:10.5pt;footnotesize:9pt
\begin{align}	
	P_t = &\frac{R^4}{C_{\text{st}}\cdot \sigma \cdot p_{\text{DC}}}\cdot \left( \frac{\delta {a}_{\text{SRL}}\pi }{\lambda e^{-1}} \right) ^{-2\left( 2K-1 \right)}\cdot \frac{1}{T_{\text{CPI}}} \notag \\
	&\cdot \operatorname{acos}^{2-4K}\left( \frac{T_{\text{CPI}}^{2}{{d}_{1}}+{{T}_{\text{CPI}}}{{d}_{2}}+{{d}_{3}}}{\sqrt{T_{\text{CPI}}^{4}{{d}_{4}}+T_{\text{CPI}}^{3}{{d}_{5}}+T_{\text{CPI}}^{2}{{d}_{6}}+{{T}_{\text{CPI}}}{{d}_{7}}+{{d}_{8}}}} \right)
	\label{equ:PtwithTconequality}
\end{align}
This formula also establishes the equivalent relationship between the three influencing factors.

\subsection{Parameters Limits} \label{subsection:parameterlimittheory}

Within a given scenario, the target's specific imaging requirements may impose constraints on the influencing factors. This section analyzes these constraints in the context of radar system implementations and ISAR imaging scenarios.

Firstly, both $P_{\text{av}}$ and  $p_{\text{DC}}$ are constrained by the system's hardware, which establishes a maximum achievable PSNR for scatterers, denoted as ${{\rho }_{\text{PSNR}}}_{\max }$. 
Additionally, the sub-procedures of super-resolution imaging algorithm, such as range alignment and phase autofocus, impose a minimum requirement for PSNR of scatterers, denoted as ${{\rho }_{\text{PSNR}}}_{\min }$.
Then ${\rho }_{\text{PSNR}\max}$ and ${\rho }_{\text{PSNR}\min}$ determine the allowable variation of $\theta_{\Delta}$ as:
\begin{align}
		\frac{{{\lambda }}}{2\delta {{a}_{\text{SRL}}}}\cdot \frac{2{{e}^{-1}}}{\pi }{{\left( \frac{1}{{{\rho }_{\text{PSNR}\max}}} \right)}^{\frac{1}{4K-2}}}\le \theta_{\Delta} \le \frac{{{\lambda }}}{2\delta {{a}_{\text{SRL}}}}\cdot \frac{2{{e}^{-1}}}{\pi }{{\left( \frac{1}{{{\rho }_{\text{PSNR}\min }}} \right)}^{\frac{1}{4K-2}}}
		\label{equ:thetaDeltalimits}
	\end{align}

The lower bound in (\ref{equ:thetaDeltalimits}) represents the minimum required $\theta_{\Delta}$  to achieve the desired imaging resolution. Given the limited observation perspective of the radar's LOS relative to synchronous orbit targets, this lower bound can assess the feasibility of ground-based radar imaging for such targets. The upper bound of  $\theta_{\Delta}$, denoted as $\theta_{\Delta_{\max}}$, stands for the maximum value below which the super-resolution imaging algorithm can operate effectively. When $\theta_{\Delta}$  exceeds this upper bound, the efficiency of the ISAR imaging system declines, even if the desired imaging resolution is surpassed.
Similarly, the constraint for $P_t$, which defines the minimum transmitted power required for the radar system,  can be formulated as:

\begin{align}
     {{P}_{t}}\ge \max \left( \frac{{{R}^{4}}}{{{C}_{\text{st}}}\cdot \sigma \cdot {{p}_{\text{DC}}}\cdot {{T}_{\text{CPI}}}\cdot {{\left( \frac{\delta {{a}_{\text{SRL}}}\cdot {{\theta }_{\Delta}}\cdot \pi }{\lambda \cdot {{e}^{-1}}} \right)}^{4K-2}}},\frac{{{R}^{4}}}{{{C}_{\text{st}}}\cdot \sigma \cdot {{p}_{\text{DC}}}\cdot {{T}_{\text{CPI}}}\cdot {{\rho }_{{{_{\text{PSNR}\min}}}}}} \right)
		\label{equ:ptminlimit}
\end{align}

It is important to note that in a specific imaging scenario, $\theta_{\Delta_{\max}}$ is influenced by multiple factors. not just the ${{\rho }_{\text{PSNR}}}_{\min }$. For High Earth Orbit (HEO) targets, $\theta_{\Delta_{\max}}$  is determined by the maximum variation of the radar LOS relative to the target. For Low Earth Orbit (LEO) targets, it reflects the maximum angle achievable under constant target scattering characteristics and algorithm limitation of the imaging algorithm. Additionally, the data rate of independent imaging frames constrains the maximum $T_{\text{CPI}}$ and thus determines $\theta_{\Delta_{\max}}$.

\begin{table*}[!b]
	\fontsize{10pt}{12pt}\selectfont
    \caption{Typical Target TLE}
    \label{table:TLEData}
    \centering
    %\tablefont
    %\multicolumn{1}{c}{COSMOS   2494}   \\   
\begin{tabular}{c}
	     COSMOS   2494 \\
	    \Xhline{1pt}
		\makebox[35em][s]{1 39491U 13078B   24183.51288981  .00004231  00000-0  37027-3 0  9997}\\% 69 characters 30em
		\makebox[35em][s]{2 39491  82.4151 229.6952 0021025 105.6371 254.7173 14.96334548 571512}\\
		\Xhline{1pt} \\
		NAVSTAR 81 \\
		\Xhline{1pt}
		\makebox[35em][s]{1 48859U 21054A   24184.56006463 -.00000094  00000-0  00000+0 0  9998}\\% 69 characters 30em
		\makebox[35em][s]{2 48859  55.3665   1.1197 0015532 218.3626 150.8174  2.00557019 22413}\\
		\Xhline{1pt} \\
		BeiDou 9 \\
		\Xhline{1pt}
		\makebox[35em][s]{1 37763U 11038A   24183.90000135  .00000035  00000-0  00000-0 0  9993}\\% 69 characters 30em
		\makebox[35em][s]{2 37763  54.5617 172.0429 0126267 229.1018 296.8774  1.00256110 47479}\\
		\Xhline{1pt} 	
\end{tabular}
\end{table*}

\section{Simulation}\label{section:simulation}

In this section, simulations are conducted to evaluate the effects and tradeoffs of the influencing factors on $\delta{a}_{\text{SRL}}$. Additionally, the observation efficiency and power constraints for super-resolution imaging of targets in different orbits are analyzed.
Three typical space targets in different orbits are selected, whose two-line orbital elements (TLE) are provided in Table \ref*{table:TLEData}. Radar system parameters are listed in Table \ref*{tabel:radarparameter}. All targets operate with Earth-oriented-3-axis stabilized (EOAS) mode for attitude control.
In the subsequent simulations, except for the specifically noted settings, all other radar parameters, target parameters, and imaging scenario settings remain consistent with this setup.

\begin{table}[H] %H表示绝对在这里
	\fontsize{10pt}{12pt}\selectfont  % 设置字体大小为10pt，行间距为12pt
	\caption{Radar System Parameters}
	\label{tabel:radarparameter}
	\centering
	\begin{tabular}{|c|c|c|}
		\hline
		parameter                             & \multicolumn{2}{c|}{value}     \\
		\hline
		\multirow{3}{*}{Radar station location} & latitude         & 46°N        \\ \cline{2-3}
		& longitude       & 130°E      \\ \cline{2-3}
		& altitude           & 0km        \\
		\hline
		radar carrier                         & \multicolumn{2}{c|}{16.7GHz}    \\
		\hline
		radar bandwidth                       & \multicolumn{2}{c|}{2GHz}       \\
		\hline
		duty cycle                      & \multicolumn{2}{c|}{20\%}       \\
		\hline
	\end{tabular}
\end{table}

\subsection{Cross-Range Resolution Contrast} \label{subsection:resolutioncontrast}

The following simulations demonstrate the relationship between the Rayleigh limit and the computational resolution limit of cross-range resolution when imaging the representative LEO space target, COSMOS 2494. The radar system records three visible passes over COSMOS 2494 within 12 hours starting from 12:00:00 on July 5, 2024. Our analysis focuses on the first pass, from 18:59:38 to 19:10:24, lasting 646 seconds. 
Let $\delta {{a}_{\text{RL}}}$ align with the range resolution, denoted by $\delta r=c/2B_w$, in ISAR image, then the azimuth cumulative rotation angle is set to $\theta_{\Delta} = {6.86}^\circ$, approximating the ratio of bandwidth carrier frequency ratio following the Rayleigh criterion.

\begin{figure}[!ht]
	\centering
	\subfloat[]{\includegraphics[width=2in]{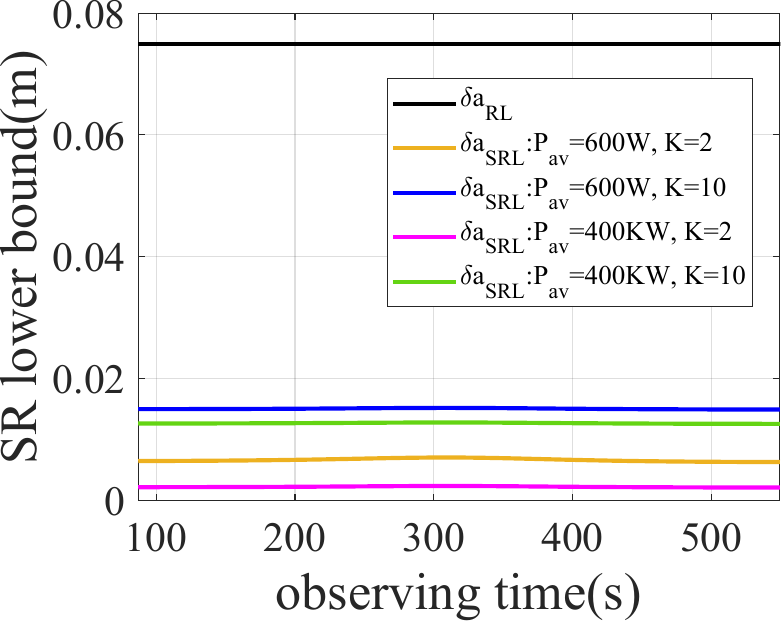}%
		\label{fig:3a}}
	\hfil
	\subfloat[]{\includegraphics[width=2in]{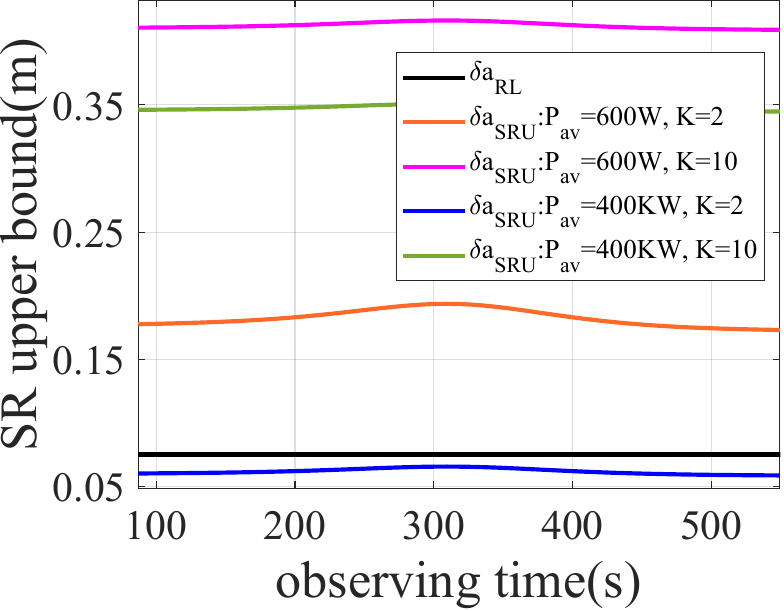}%
		\label{fig:3b}}
	\caption{the Rayleigh limit and the computational resolution limit of cross-range resolution under different $K$ and $P_t$  during the first pass. (a) $\delta{a}_{\text{RL}}$ and $\delta{a}_{\text{SRL}}$. (b) $\delta{a}_{\text{RL}}$ and $\delta{a}_{\text{SRU}}$ (the legend is consistent with (a)).}
	\label{fig:3}
\end{figure}
\begin{figure}[ht]
	\centering
	\includegraphics[width=2in]{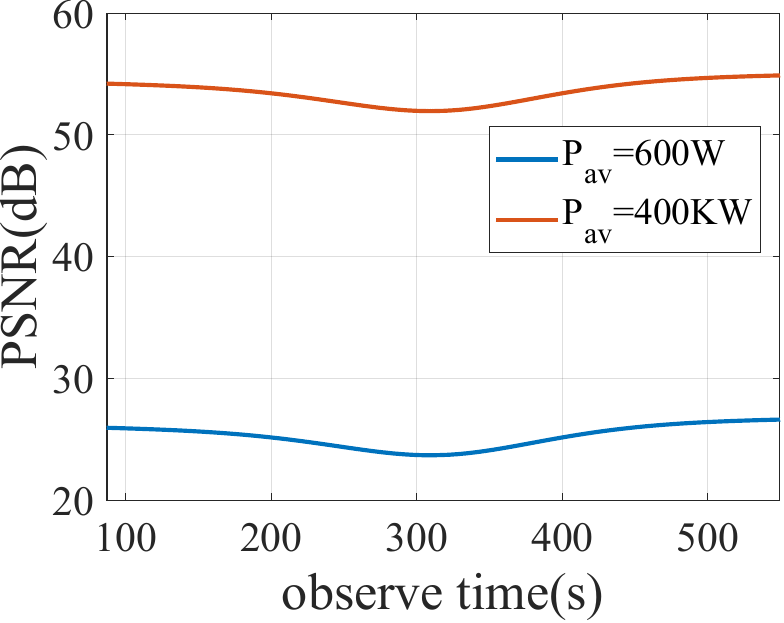}%
	\caption{the PSNR required under different $P_{\text{av}}$.}
	\label{fig:4}
\end{figure}

Fig.\ref{fig:3} illustrates the relationship between $\delta{a}_{\text{RL}}$ and the upper and lower bounds of  $\delta {{a}_{\text{cmp}}}$, as well as the influence of different average transmitted powers and the number of scatterers on these bounds. The number of scatterers $K$ in a range cell is set to 2 and 10, respectively, while the radar average transmitted power $P_{\text{av}}$ is set to 600W and 400KW, respectively. 
It can be observed that when $K=10$, $\delta{a}_{\text{RL}}$ lies between the upper and lower bounds of $\delta{a}_{\text{cmp}}$, and $\delta{a}_{\text{SRL}}$ is being significantly smaller than $\delta{a}_{\text{RL}}$. 
In imaging scenarios with a small $K$ and high PSNR (e.g., $P_{\text{av}}=400KW$, $K=2$), both the upper and lower bounds of $\delta{a}_{\text{cmp}}$ will be lower than the Rayleigh limit. 
As shown in Fig.\ref{fig:4}, the PSNR of the scatterers ranges from 50 dB to 55 dB when $P_{\text{av}}=400KW$. However, achieving such high PSNR values in real ISAR imaging scenarios is challenging, making this result more theoretical than practical.

\subsection{The Influencing Factors' Impacts on $\delta{a}_{\text{SRL}}$}

This section simulates the impacts of cumulative rotation angle, PSNR, and number of scatterers on the lower bound of the computational resolution limit. The ISAR image time is the 325th second after the beginning of the first pass of COSMOS 2494.

\begin{figure}[!h]
	\centering
	\subfloat[]{\includegraphics[width=1.6in]{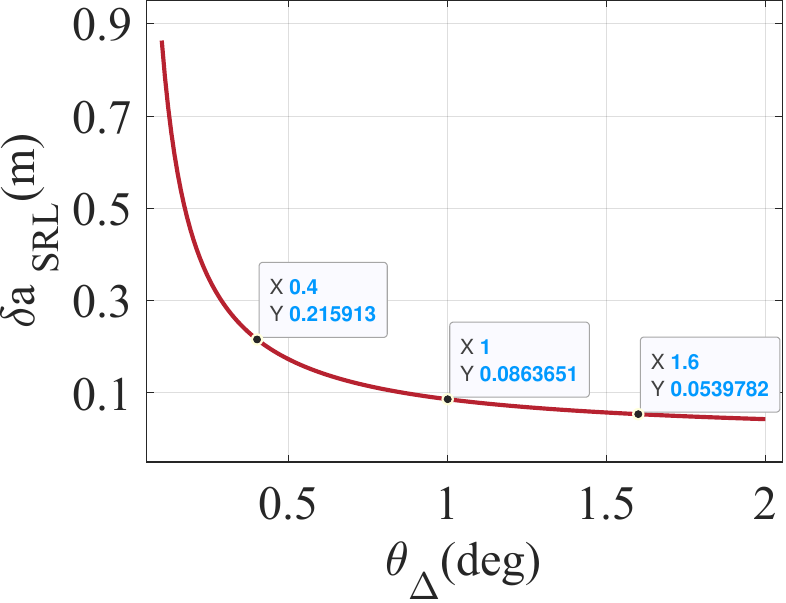}%
		\label{fig:funcSRLwiththeta}}
	\hfil
	\subfloat[]{\includegraphics[width=1.6in]{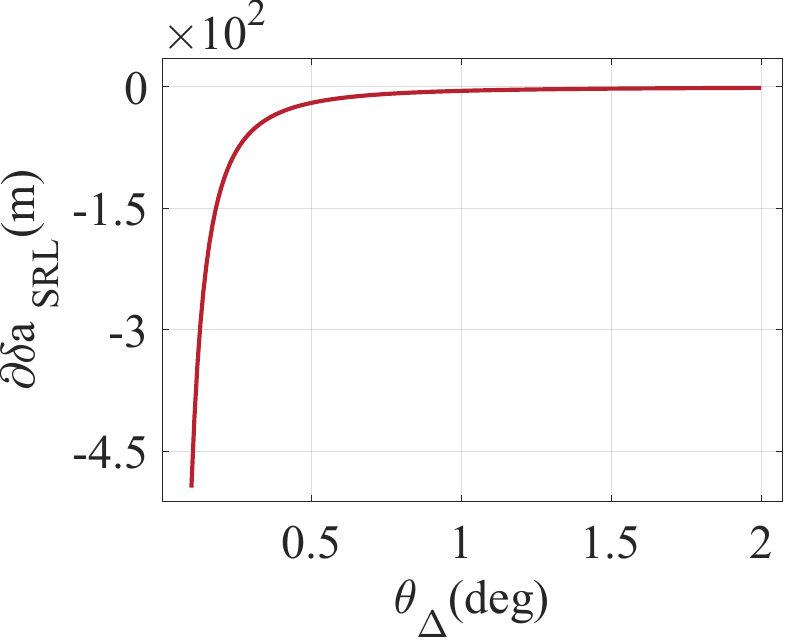}%
      \hfil
		\label{fig:derifuncSRLwiththeta}}
	\subfloat[]{\includegraphics[width=1.6in]{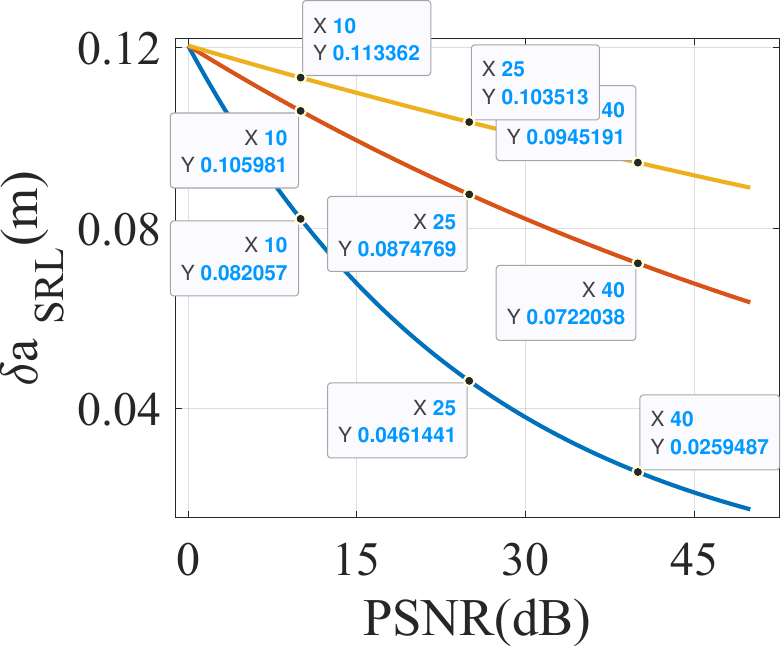}%
		\label{fig:funcSRLwithPSNR}}
	\hfil
	\subfloat[]{\includegraphics[width=1.6in]{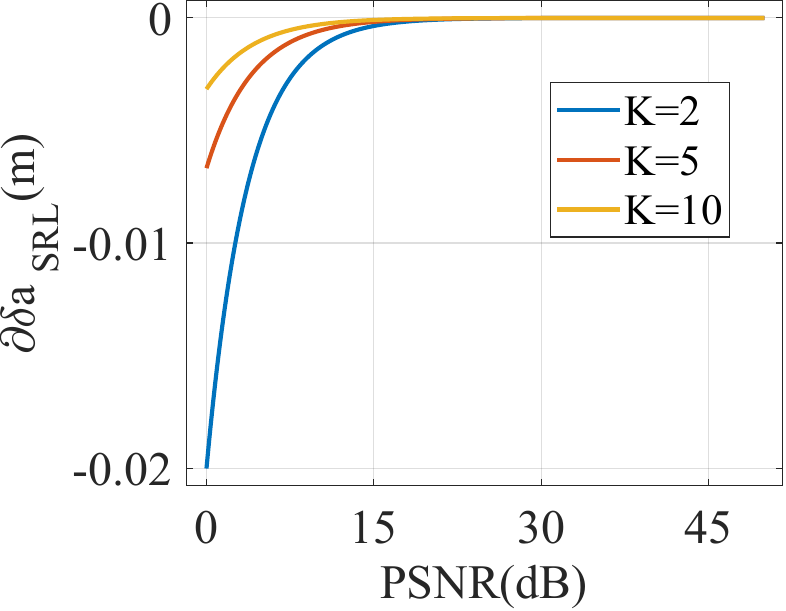}%
		\label{fig:derifuncSRLwithPSNR}}
	%\centering
     \qquad
	\subfloat[]{\includegraphics[width=1.6in]{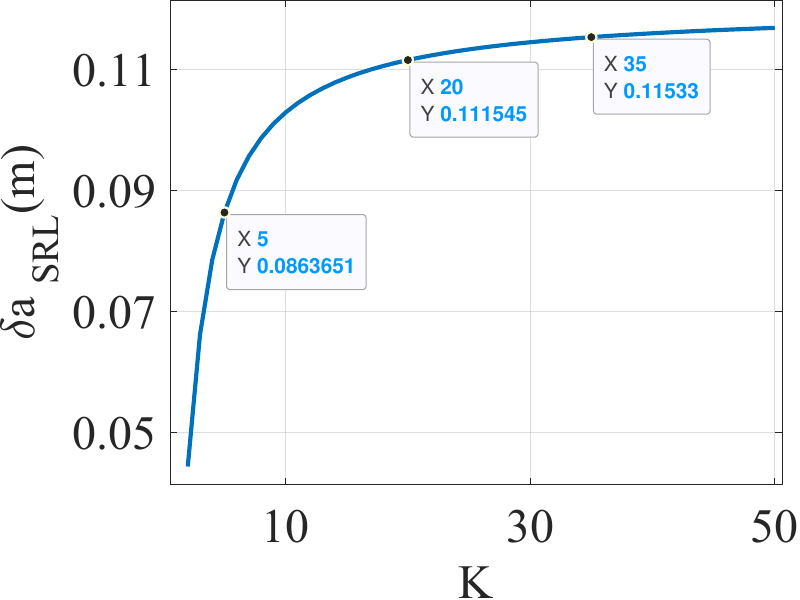}%
		\label{fig:funcSRLwithK}}
	\hfil
	\subfloat[]{\includegraphics[width=1.6in]{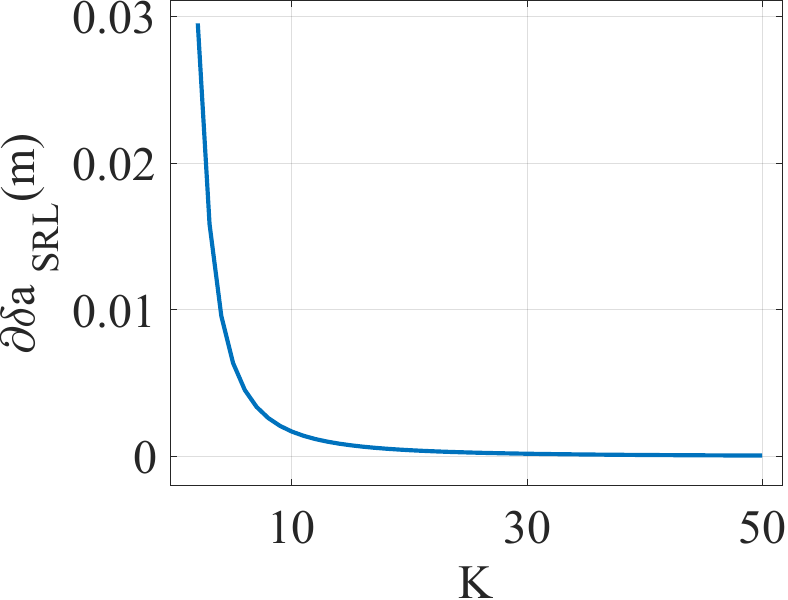}%
		\label{fig:derifuncSRLwithK}}
    \hfil
	\subfloat[]{\includegraphics[width=1.6in]{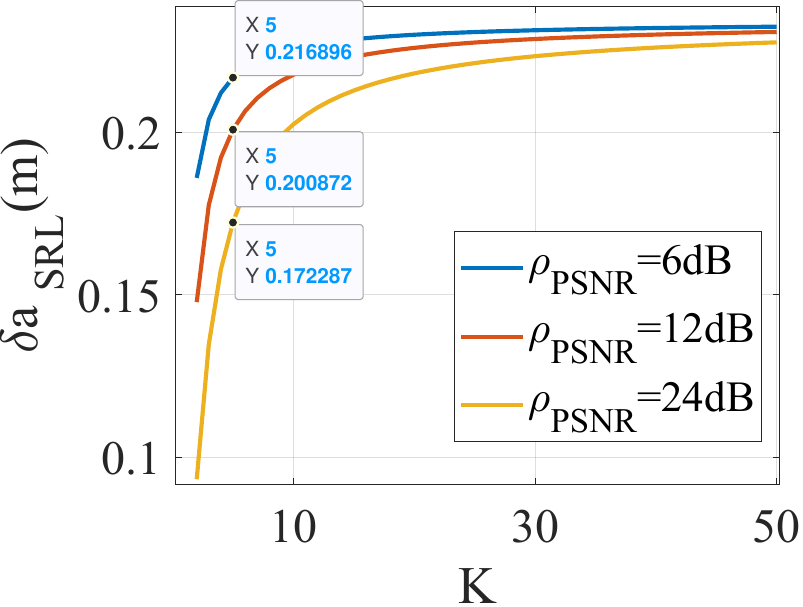}%
		\label{fig:funcSRLwithKandPSNR}}
	\hfil
	\subfloat[]{\includegraphics[width=1.6in]{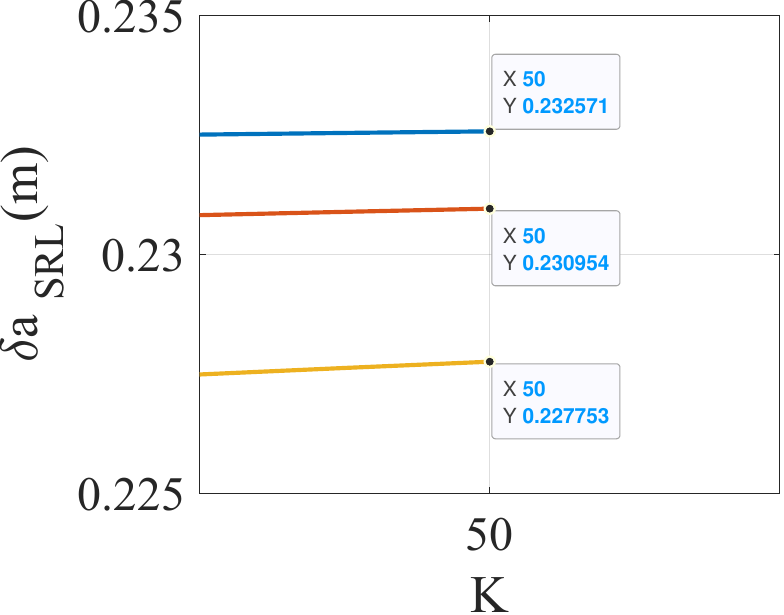}%
		\label{fig:funcSRLwithKandPSNRlocal}}
	\caption{the impacts of influencing factors on the super-resolution limit. (a)-(b) the variations of $\delta{a}_{\text{SRL}}$ and $\partial \delta {{a}_{\text{SRL}}}$ as $\theta_{\Delta}$ increases when $\rho_{\text{PSNR}}=26$dB and $K=5$. (c)-(d) the variations of $\delta{a}_{\text{SRL}}$ and $\partial \delta {{a}_{\text{SRL}}}$ as $\rho_{\text{PSNR}} $ increases when $\theta_{\Delta}=0.5^{\circ}$ and $K=5$. (e)-(f) the variations of $\delta{a}_{\text{SRL}}$ and $\partial \delta {{a}_{\text{SRL}}}$ as $K$ increases when $\rho_{\text{PSNR}}=26$dB and $\theta_{\Delta}=0.5^{\circ}$. (g)-(h) the constringency  of  the super-resolution limit and $K$ with different $\rho_{\text{PSNR}}$.}
	
	%\caption{the impact of influencing factors on the super-resolution limit $\delta{a}_{\text{SRL}}$ when $\rho_{\text{PSNR}}=26dB$ and $\theta_{\Delta}=0.5^{\circ}$. (a) the variations of $\delta{a}_{\text{SRL}}$ as $K$ changes. (b) the variations of $\partial \delta {{a}_{\text{SRL}}}$ as $K$ changes.}
	\label{fig:theimpactofinfluencingfactors}
\end{figure}

Fig.\ref{fig:funcSRLwiththeta} and Fig.\ref{fig:derifuncSRLwiththeta} illustrate the impact of $\theta_{\Delta} $ on $\delta{a}_{\text{SRL}}$ when $\rho_{\text{PSNR}}=26\text{dB}$ and $K=5$.
In Fig.\ref{fig:funcSRLwiththeta}, $\delta a_{\text{SRL}}$ is 0.216$m$ at $\theta_{\Delta} ={{0.4}^\circ}$. As $\theta_{\Delta}$ increases to $ {{1.0}^\circ}$, $\delta a_{\text{SRL}}$ decreases to 0.086$m$, marking a 60.00$\%$ reduction of $\delta a_{\text{SRL}}$ compared to $\theta_{\Delta}={{0.4}^\circ}$. Further increasing $\theta_{\Delta}$ to $ {{1.6}^\circ}$ reduces $\delta a_{\text{SRL}}$ to 0.054$m$, indicating a 37.2$\%$ decrease of $\delta a_{\text{SRL}}$ relative to when $\theta_{\Delta}={{1.0}^\circ}$ . 
In Fig.\ref{fig:derifuncSRLwiththeta}, when $\theta_{\Delta}$ exceeds $1^\circ$, the improvement in super-resolution limit from further increases in
$\theta_{\Delta}$ gradually diminishes.
These results align with the conclusions in Section \ref{section:influencinfactors}, highlighting the significance of the marginal income from the $\theta_{\Delta}$  in imaging scenario design.

Fig.\ref{fig:funcSRLwithPSNR} and Fig.\ref{fig:derifuncSRLwithPSNR} illustrate the impact of $\rho_{\text{PSNR}} $ on $\delta{a}_{\text{SRL}}$ when $\theta_{\Delta}=0.5^{\circ}$. Table \ref{table:figdata} summarizes the selected results for different cases shown in  Fig.\ref{fig:funcSRLwithPSNR}, where the percentages in the last column represent the reduction in $\delta {{a}_{\text{SRL}}}$ compared to the previous row's parameter settings. Additionally, as shown in Fig.\ref{fig:derifuncSRLwithPSNR}, the improvement rate of $\delta a_{\text{SRL}}$ diminishes as ${\rho }_{\text{PSNR}} $ increases, demonstrating that enhancing ${\rho }_{\text{PSNR}} $ by an order of magnitude approximately results in a linear reduction of $\delta {{a}_{\text{SRL}}}$. Moreover, the rate of improvement due to an increase in ${\rho }_{\text{PSNR}}$ is also influenced by $K$.

Fig.\ref{fig:funcSRLwithK} and Fig.\ref{fig:derifuncSRLwithK} illustrate the impact of scatterer numbers on the super-resolution limit when $\rho_{\text{PSNR}}=26\text{dB}$ and $\theta_{\Delta}=0.5^{\circ}$.
In Fig.\ref{fig:funcSRLwithK},  $\delta a_{\text{SRL}}$ is 0.086$m$ at $K=5$. As $K$ increases to 20, $\delta a_{\text{SRL}}$ rises to 0.112$m$, reflecting a 29.2$\%$ increase compared to $K=5$. At $K=35$, $\delta a_{\text{SRL}}$ further increases to 0.115$m$, representing only 3.4$\%$ growth relative to $K=20$.
In Fig.\ref{fig:derifuncSRLwithK}, as $K$ exceeds 20, each subsequent increase in $K$ leads to a progressively smaller decrease in $\delta a_{\text{SRL}}$, indicating a nonlinear relationship between $K$ and $\delta a_{\text{SRL}}$.  Initially, increasing $K$ significantly degrades imaging performance, but as $K$ continues to grow, its negative impact on super-resolution imaging performance gradually diminishes.

\vspace{0.5cm}
\begin{minipage}[!t]{0.5\textwidth}
\centering
\fontsize{10pt}{10pt}\selectfont  % 设置字体大小为10pt，行间距为12pt
\captionof{table}{Data in Fig.\ref{fig:funcSRLwithPSNR}}
\label{table:figdata}
\begin{tabular}{cccc}
		\Xhline{1.5pt}
		\makecell{Scatterer \\ Numbers}   & PSNR(dB) & $\delta a_{\text{SRL}}$(m)   & \makecell{$\delta a_{\text{SRL}}$ \\ Reduction} \\ \Xhline{1.5pt}
		\multirow{3}{*}{K=2}  & 10       & 0.082 &  \diagbox{}{} \\ %\cline{2-4}
		& 25       & 0.046 & 43.8\%    \\ %\cline{2-4}
		& 40       & 0.026 & 43.8\%    \\ \hline
		\multirow{3}{*}{K=5}  & 10       & 0.106 & \diagbox{}{} \\ %\cline{2-4}
		& 25       & 0.087 & 17.5\%    \\ %\cline{2-4}
		& 40       & 0.072 & 17.5\%    \\ \hline
		\multirow{3}{*}{K=10} & 10       & 0.113 & \diagbox{}{} \\ %\cline{2-4}
		& 25       & 0.104 & 8.69\%    \\ %\cline{2-4}
		& 40       & 0.095 & 8.69\%    \\ \Xhline{1.5pt}
\end{tabular}
\end{minipage}
\begin{minipage}[c]{0.5\textwidth}
\centering
\fontsize{10pt}{12pt}\selectfont  % 设置字体大小为10pt，行间距为12pt
\captionof{table}{Data in Fig.\ref{fig:funcSRLwithKandPSNR}}
\label{table:dataratio}
\begin{tabular}{cccc}
		\Xhline{1.5pt}
		\makecell{Scatterer \\ Numbers}
		& PSNR (dB) & ratio (\%) & \makecell{Resolution \\ Degrade} \\ \Xhline{1.5pt}
		\multirow{3}{*}{K=5} 
		& 6 & 21.69 &  \diagbox{}{} \\ %\cline{2-4}
		& 12 & 20.09 & 7.37\% \\ %\cline{2-4}
		& 24 & 17.23 & 14.24\% \\ \hline
		\multirow{3}{*}{K=50} 
		& 6 & 23.26 &  \diagbox{}{} \\ %\cline{2-4}
		& 12 & 23.10 & 0.69\% \\ %\cline{2-4}
		& 24 & 22.78 & 1.39\% \\ \Xhline{1.5pt}
\end{tabular}
\end{minipage}
\vspace{0.5cm}

Fig.\ref{fig:funcSRLwithKandPSNR} and Fig.\ref{fig:funcSRLwithKandPSNRlocal} show the radio of the super-resolution limit to the Rayleigh limit under different ${\rho }_{\text{PSNR}}$ as $K$ increases with a fixed $\theta_{\Delta }$. Table \ref*{table:dataratio} presents the selected results from Fig.\ref{fig:funcSRLwithKandPSNR} and Fig.\ref{fig:funcSRLwithKandPSNRlocal}.
According to Fig.\ref{fig:funcSRLwithKandPSNR} and Table \ref*{table:dataratio}, it is evident that the impact of the same increment of ${\rho }_{\text{PSNR}}$  on $\delta a_{\text{SRL}}$  becomes progressively weaker as $K$  increases, causing the super-resolution limit converges to $2e^{-1}/\pi \approx 0.234$ times the Rayleigh limit. 
This demonstrates the crucial role of $K$  in determining the performance of super-resolution algorithms. The superior performance of super-resolution algorithms is primarily due to scatterer sparsity, though this does not mean that $K \ll N$ ( where $N$ is the sampling number in the azimuth direction of ISAR image). Instead, it reflects the relatively small value of $K$.  When the value of $K$  is large, if super-resolution algorithms perform well with larger $K/N$, this is mainly due to the benefit of high $\rho_{\text{PSNR}}$ achieved through the coherent integration as $N$ increases.

\subsection{Parameters Limits} \label{subsection:parameterslimit}

The following simulations focus on the minimum requirements for $\rho_{\text{PSNR}}$, $P_{\text{av}}$, and $\theta_{\Delta}$ under constant transmitted energy to achieve a desired cross-range computational resolution limit. In these simulations, $K$ is set to 10, and the desired cross-range super-resolution limit is aligned with the range resolution, meaning $\delta {{a}_{\text{SRL}}}=\delta r$. The ISAR imaging time is the first pass of COSMOS 2494, from 18:59:38 to 19:10:24.

\begin{figure*}[!h] %t:top，顶部；b：bottom，底部；h：here，当前位置;p:page,浮动页。!:忽视大部分对浮动体位置的内建限制，尽可能地满足我们的位置要求
	\centering
	\subfloat[]{\includegraphics[width=1.9in]{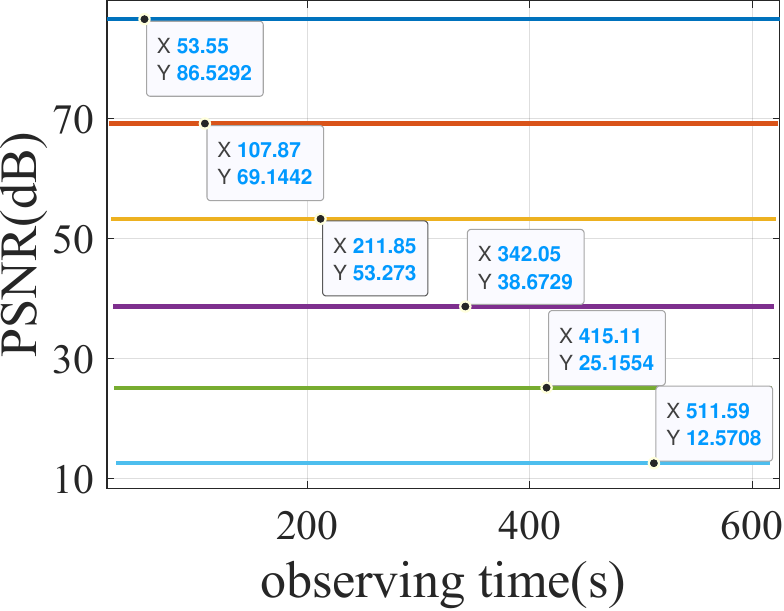}%
		\label{fig:PSNR}}
	\hfil
	\subfloat[]{\includegraphics[width=1.9in]{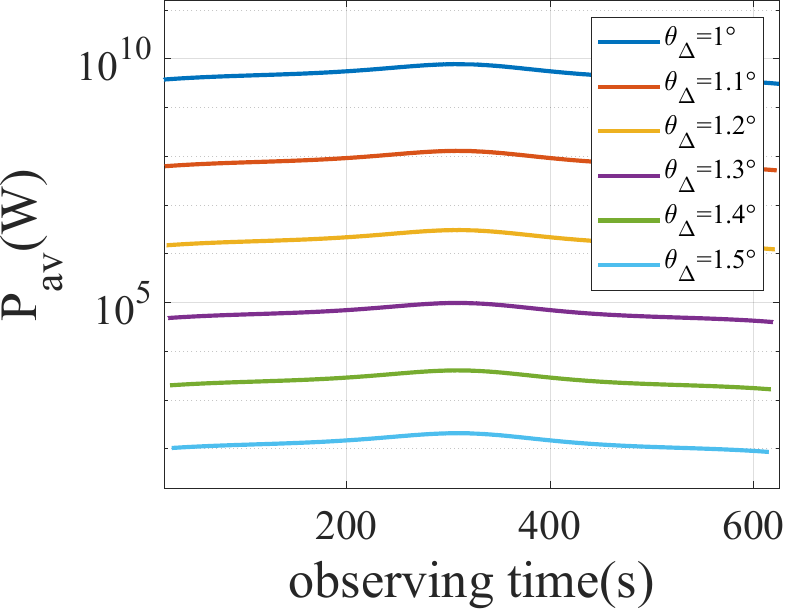}%
		\label{fig:Pav}}
	\hfil
	\subfloat[]{\includegraphics[width=1.9in]{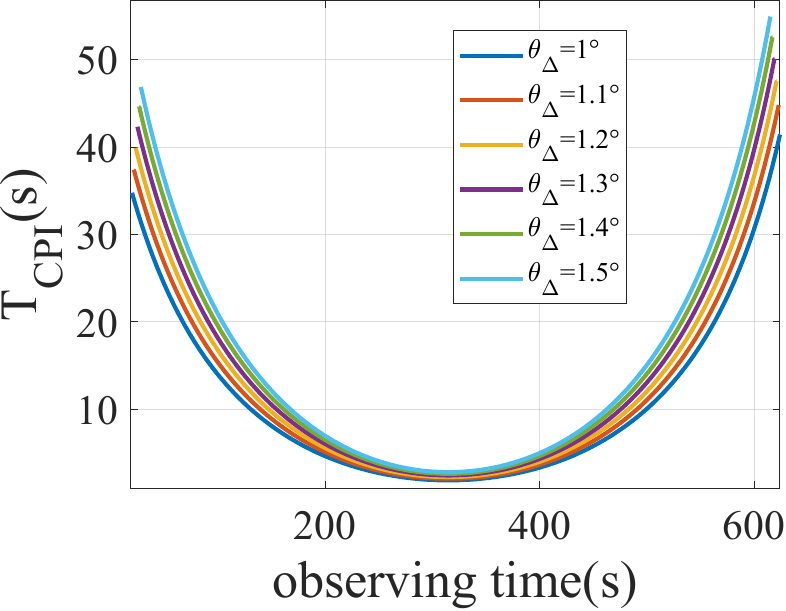}%
		\label{fig:Tcon}}
	\caption{the minimum requirements for $\rho_{\text{PSNR}}$, ${P_\text{av}}$, and $T_{\text{CPI}}$ under different $\theta_{\Delta}$. (a) the required $\rho_{\text{PSNR}}$ (the legend is consistent with (b)). (b) the required ${P_\text{av}}$. (c) the required $T_{\text{CPI}}$ .}
	\label{fig:PSNRandPav}
\end{figure*}

As described in (\ref{equ:PSNRfucntion}), the minimum requirement for $\rho_{\text{PSNR}}$ or ${P_\text{av}}$ to achieve the desired $\delta{a}_{\text{SRL}}$ can be determined based on the relationship between $\rho_{\text{PSNR}}$ and ${P_\text{av}}$ under different $\theta_{\Delta}$. Fig.\ref{fig:PSNRandPav} shows the minimum requirements for $\rho_{\text{PSNR}}$, ${P_\text{av}}$, and $T_{\text{CPI}}$ across these varying  $\theta_{\Delta}$.
As shown in Fig.\ref{fig:PSNR}, a smaller $\theta_{\Delta}$ corresponds to a much higher $\rho_{\text{PSNR}}$, this nonlinear relationship satisfies $\rho_{\text{PSNR}} \ge {\left({c\pi \theta_{\Delta} }/{2B_w \lambda e^{-1}}\right) }^{-4K-2}$, which can be derived from (\ref{equ:crossrangeresolutionSRL}).
Consequently, the minimum required $\rho_{\text{PSNR}}$ for achieving the desired $\delta{a}_{\text{SRL}}$ remains constant at different imaging time if both $K$ and $\theta_{\Delta}$ are unchanged.  
Additionally, the fluctuation of ${P_\text{av}}$ under a specific $\theta_{\Delta}$, as shown in Fig.\ref{fig:Pav}, indicates that the minimum requirement for $P_{\text{av}}$ varies with the target’s effective rotation rate and the round-trip distance between the target and the radar.

Specifically, as shown in Fig.\ref{fig:PSNRandPav}, when $\theta_{\Delta}$ falls below ${1.2}^\circ$, the minimum required $\rho_{\text{PSNR}}$ is 53.27dB, with the corresponding minimum required $P_{\text{av}}$ ranging from 1MW to 10MW, imposing a significant burden on the radar system. 
Fortunately, extending the $T_{\text{CPI}}$ can effectively increase $\theta_{\Delta}$, thereby reducing the required $P_{\text{av}}$. 
A comparison of the results in Fig.\ref{fig:PSNRandPav} reveals that when $\theta_{\Delta}$ is between ${1.3}^\circ$ and ${1.5}^\circ$, the required minimum  $T_{\text{CPI}}$ to obtain an ISAR image is within the 60s, while the required $P_{\text{av}}$ remains below 100KW, which is more feasible for a operational radar system. 

Note that these simulations provide a representative example for the minimum requirements for $P_{\text{av}}$ and $\rho_{\text{PSNR}}$ under different $\theta_{\Delta}$. However, the minimum requirements in other imaging scenarios should be determined based on the specific imaging targets and radar system, rather than directly applying these results. Moreover, their requirements should not exceed the limits discussed in Section \ref{subsection:parameterlimittheory}.

\subsection{The Tradeoff between $p_{\text{DC}}$ and $T_{\text{CPI}}$ with Constant $E$}

The following simulations focus on the tradeoff between $p_{\text{DC}}$ and  $T_{\text{CPI}}$ (changes with $\theta_{\Delta}$) with constant $E$ in super-resolution imaging, and its impact on super-resolution limit. The ISAR imaging time is the first pass of COSMOS 2494, from 18:59:38 to 19:10:24.

\begin{figure*}[!ht]
	\centering
	\subfloat[]{\includegraphics[height=1.2in]{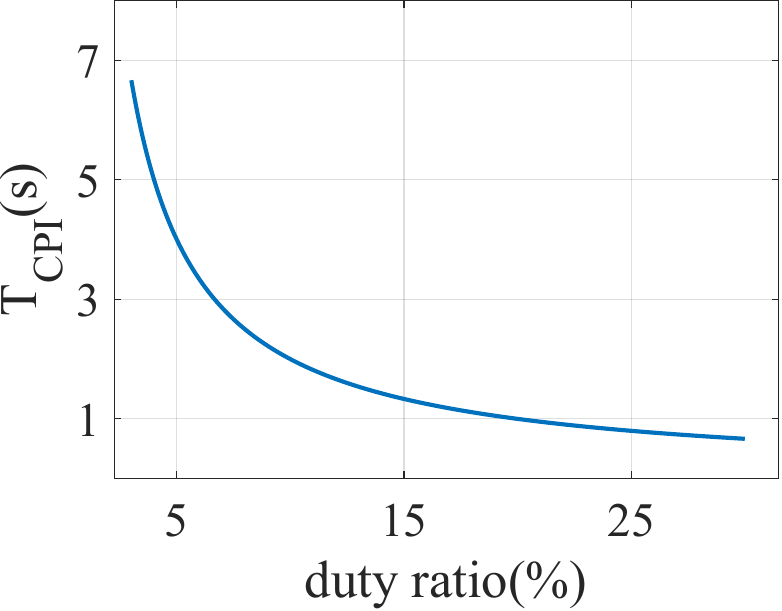}%
		\label{fig:TconwithpDR}}
	\hfil
	\subfloat[]{\includegraphics[height=1.2in]{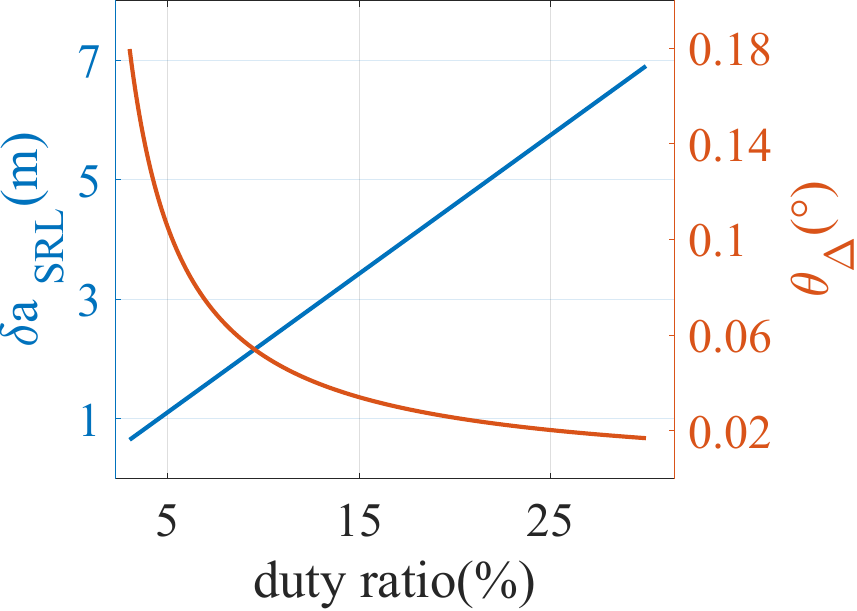}%
		\label{fig:SRLandthetawithpDR}}
	\hfil
	\subfloat[]{\includegraphics[height=1.2in]{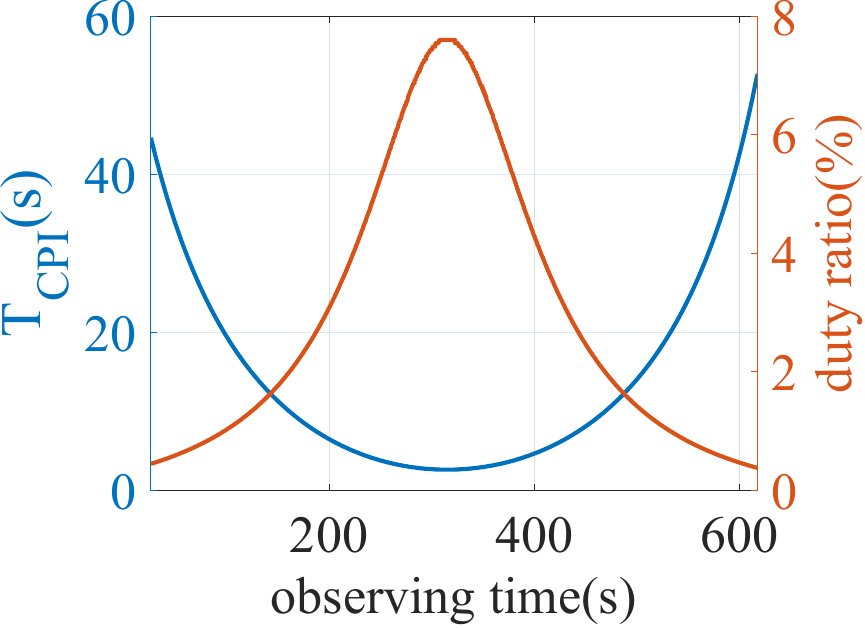}%
		\label{fig:Tconwithtime}}
	\hfil
	\subfloat[]{\includegraphics[height=1.2in]{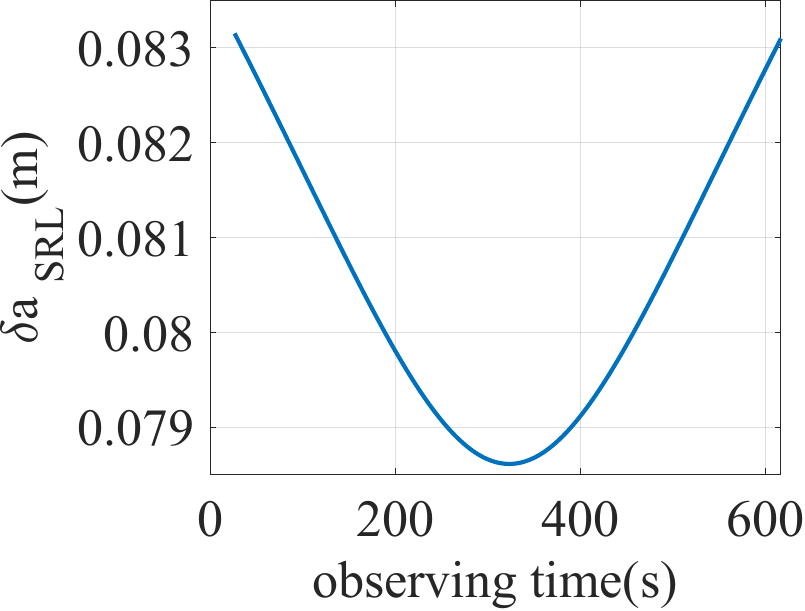}%
		\label{fig:SRLwithtime}}
	\caption{the tradeoff between $p_{\text{DC}}$ and $T_{\text{CPI}}$ with constant $E$ and its impact on $\delta a_{\text{SRL}}$. (a) the impact of $p_{\text{DC}}$ on $T_{\text{CPI}}$ with constant $E$. (b) the impact of $p_{\text{DC}}$ on $\theta_{\Delta}$ and $\delta a_{\text{SRL}}$ with constant $E$. (c) the requirements for $T_{\text{CPI}}$ and $p_{\text{DC}}$ when $\theta_{\Delta} = 1.4^{\circ}$ and $E$ holds constant. (d) the $\theta_{\Delta}$ when $\theta_{\Delta} = 1.4^{\circ}$ and $E$ holds constant. }
\end{figure*}

Fig.\ref{fig:TconwithpDR} illustrates the tradeoff between $p_{\text{DC}}$ and $T_{\text{CPI}}$ for constant $E$ when $P_{\text{av}} = 1KW$, as shown in eq.(\ref{equ:TconwithE}). Fig.\ref{fig:SRLandthetawithpDR}  shows the variations of $\delta{a}_{\text{SRL}}$ as $\theta_{\Delta}$ increases with the fixed $E$, as shown in eq.(\ref{equ:SRLwithTconandE}). It's evident that as $p_{\text{DC}}$ increases,  $\theta_{\Delta}$ decreases with $T_{\text{CPI}}$, resulting in a continuous decline in the super-resolution capability of the ISAR imaging system.  When $p_{\text{DC}}$ increases from 3$\%$ to 30$\%$ with constant $E$, the super-resolution limit degrades from 0.65m to 6.90m.

Fig.\ref{fig:Tconwithtime} shows the requirements for $T_{\text{CPI}}$ and $p_{\text{DC}}$ when $\theta_{\Delta}={1.4}^\circ$ and E holds constant, and the impacts of these requirements on the super-resolution limit is depicted in Fig.\ref{fig:SRLwithtime}. 
Although the $\rho_{\text{PSNR}}$ is lower during the rise and set periods compared to the zenith time due to the round-trip distance \cite{yuan2018phase}, Fig.\ref{fig:SRLwithtime} demonstrates that by adjusting $T_{\text{CPI}}$ and $p_{\text{DC}}$, the super-resolution capability can be effectively maintained with minimal variation across the rise, set, and zenith times. The changes in $\delta{a}_{\text{SRL}}$ are small enough to be considered negligible, allowing the super-resolution capability to be regarded as consistent throughout.

It should be noted that, without causing azimuth ambiguity, it is preferable to reduce $p_{\text{DC}}$  by lowering PRF rather than the pulse width. A larger pulse width ensures a higher $\rho_{\text{PSNR}}$  of scatterers in a single pulse, thereby reducing the system's coherence requirements for signal processing algorithms. Moreover, using fewer pulses decreases the computational burden of subsequent imaging processing.

For a multifunctional array radar with instantaneous beam-switching capability, this tradeoff allows dynamic scheduling of radar waveforms based on the imaging scenario when performing ISAR imaging of multiple targets. This approach enables the radar to maintain the required imaging resolution for each target while significantly enhancing the overall imaging efficiency of the radar system.

\subsection{The Tradeoff between $T_{\text{CPI}}$ and $P_{\text{av}}$ under different orbits}

According to Kepler's laws of planetary motion, a higher orbital altitude results in a lower relative angular velocity of the space target. Consequently, a short $T_{\text{CPI}}$  can meet the $\theta_{\Delta}$  requirement for cross-range high-resolution ISAR imaging of a LEO target.  Conversely, for HEO targets, such as the Medium Earth Orbit (MEO) or Inclined Geosynchronous Orbit (IGSO) satellites, a longer $T_{\text{CPI}}$ is needed to obtain the required $\theta_{\Delta}$. In some cases, obtaining the necessary $\theta_{\Delta}$  may not be feasible. Increasing $P_{\text{av}}$  is the only option to reduce the $\theta_{\Delta}$ requirement, as shown in equation (\ref{equ:ptminlimit}). 

In this section, several simulations for LEO, MEO, and IGSO satellites are performed to analyze the requirements for $T_{\text{CPI}}$ and $P_{\text{av}}$, as well as the tradeoffs between them, to achieve the desired cross-range resolution under different orbital altitudes and visible passes. In these simulations, the minimum $\rho_{\text{PSNR}}$  is set to 24dB, the $K$ is set to 10, the cross-range resolution is specified as $\delta{a}_{\text{SRL}}=\delta{r}$. Under these conditions, the required $\theta_{\Delta}$ for the desired super-resolution limit is calculated to be ${1.39}^\circ$.

\vspace{0.1cm}
\noindent {\bf{LEO satellite: COSMOS 2494}}
\vspace{0.1cm}

As analyzed in Section \ref{subsection:resolutioncontrast}, there are three visible passes in a 12-hour observation of the LEO satellite COSMOS 2494. Due to differences in visual range and target rotation rate across each visible arc, the requirements for $P_{\text{av}}$ and $T_{\text{CPI}}$ vary accordingly.

\begin{figure}[!t] %t:top，顶部；b：bottom，底部；h：here，当前位置;p:page,浮动页。!:忽视大部分对浮动体位置的内建限制，尽可能地满足我们的位置要求
	\centering
	\subfloat[]{\includegraphics[width=2in]{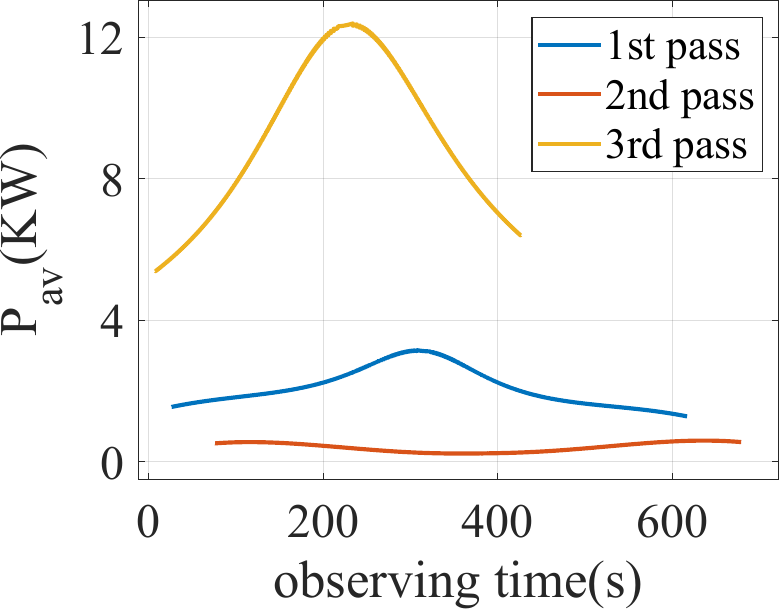}%
		\label{fig:pavofCOSMOS}}
	\hfil
	\subfloat[]{\includegraphics[width=2in]{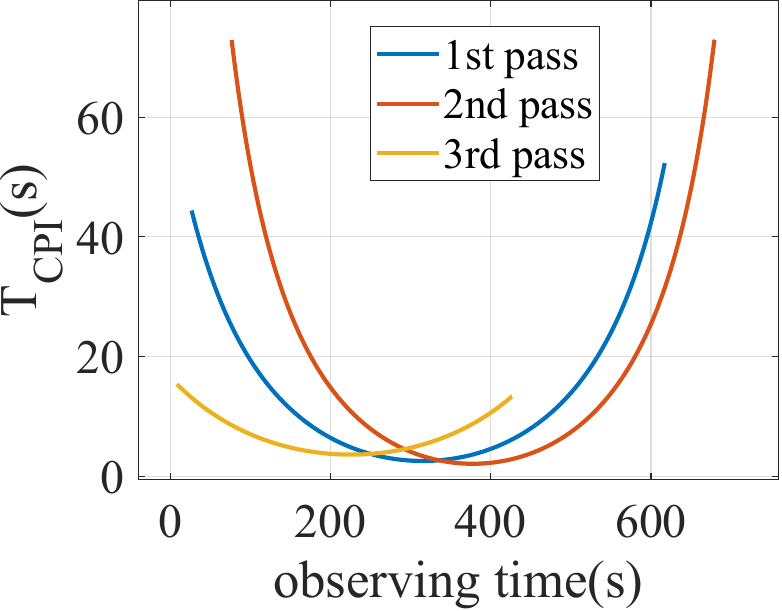}%
		\label{fig:TconofCOSMOS}}
	\caption{The minimum requirement for $P_{\text{av}}$ and $T_{\text{CPI}}$ in different visible passes of COSMOS 2494. (a) The required $P_{\text{av}}$. (b) The required $T_{\text{CPI}}$.}
	\label{fig:TconandPavofCOSMOS}
	
	\vspace{0.2cm} % 在两部分之间增加垂直间距

	\subfloat[]{\includegraphics[width=2.5in]{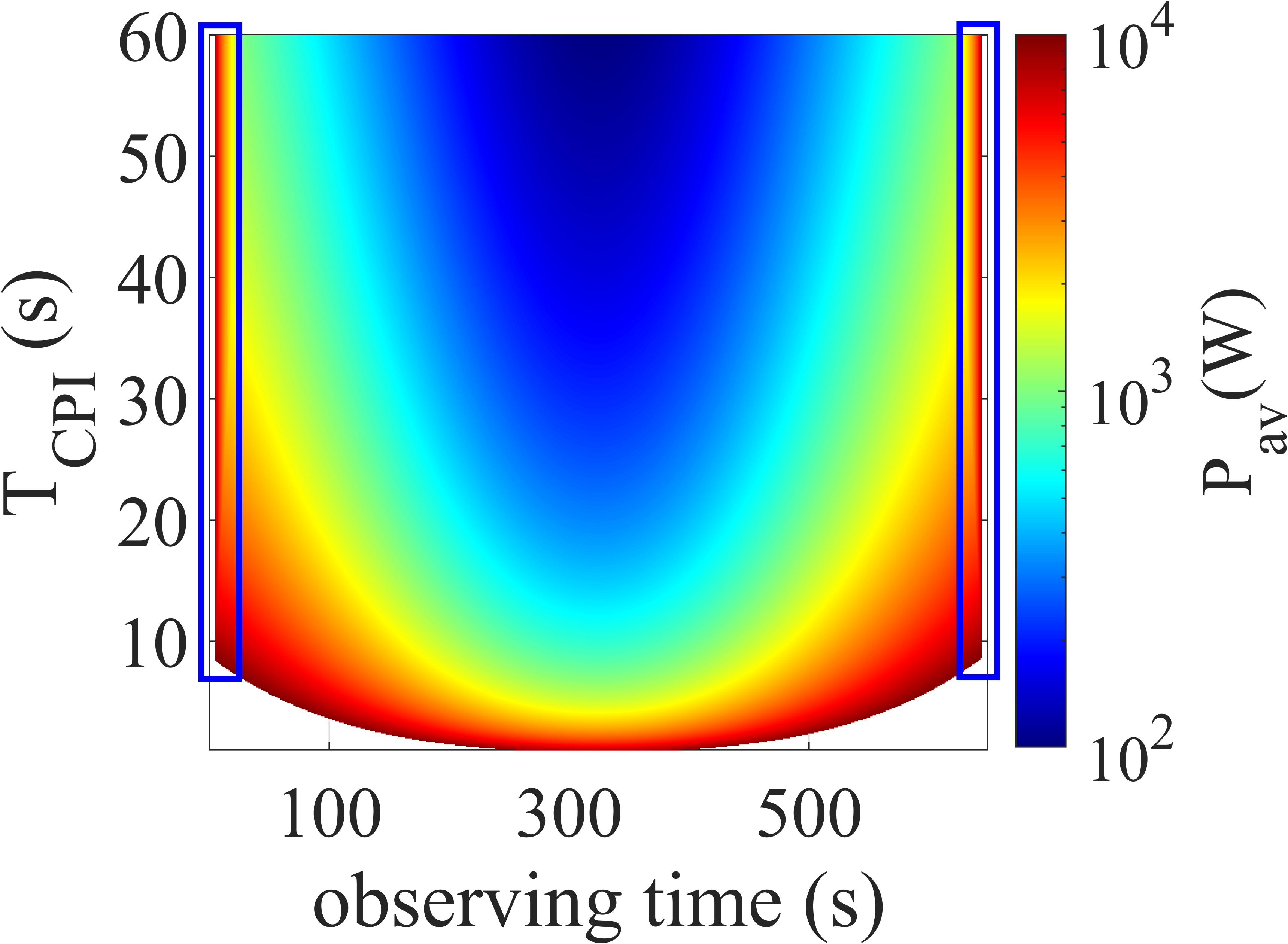}%
		\label{fig:figPavtotalLEO}}
	\hfil
	\subfloat[]{\includegraphics[width=2.5in]{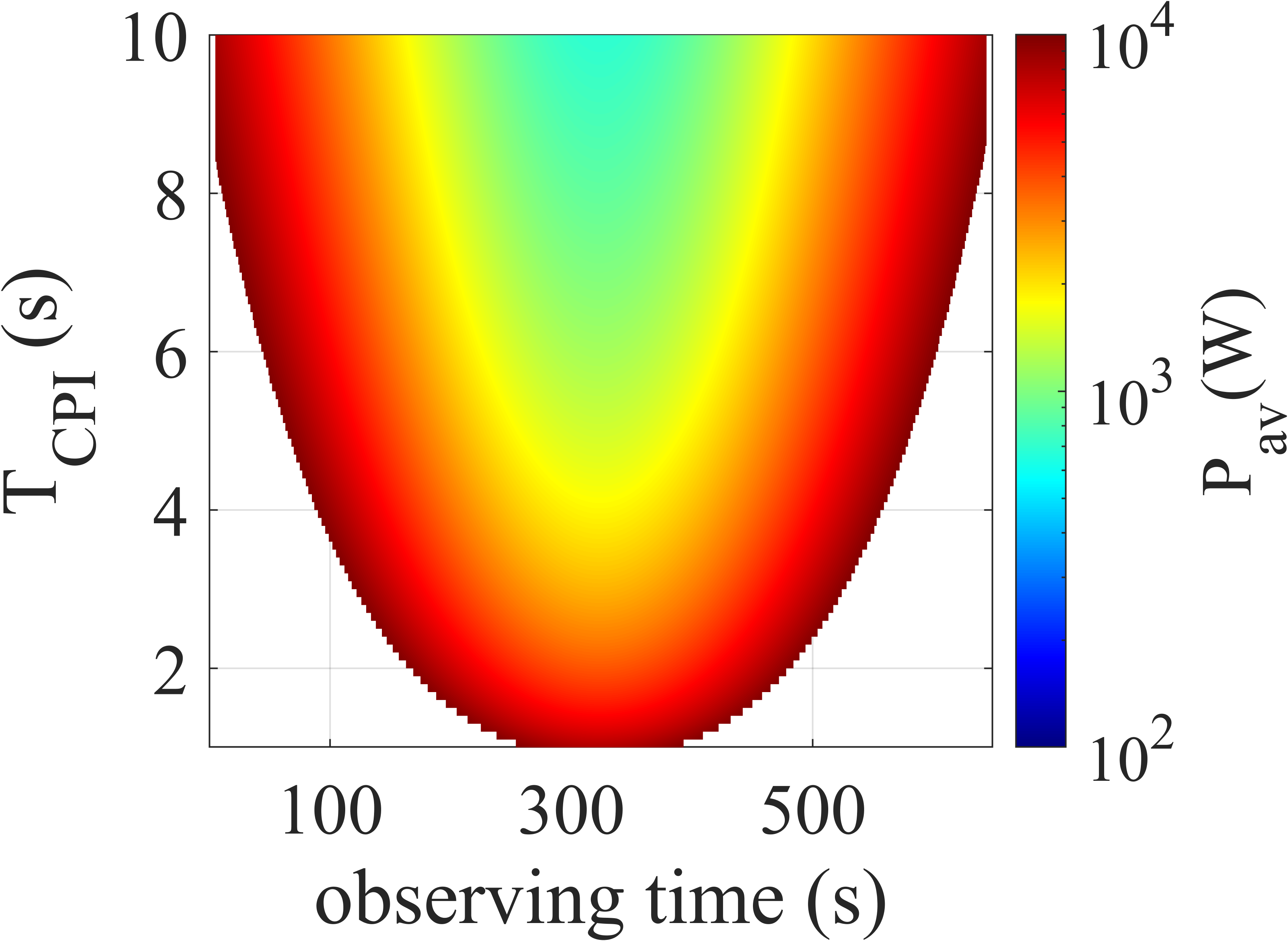}%
		\label{fig:figPavlocalLEO}}
	\caption{The tradeoff between $P_{\text{av}}$ and $T_{\text{CPI}}$ with constant $\delta{a}_{\text{SRL}}$  (pass 1). (a) The total $P_{\text{av}}$. (b) The partial $P_{\text{av}}$.}
	\label{fig:LEOPav}
\end{figure}

Fig.\ref{fig:TconandPavofCOSMOS} depicts the minimum requirements of $P_{\text{av}}$ and $T_{\text{CPI}}$ for imaging COSMOS 2494. According to Fig.\ref{fig:pavofCOSMOS}, the required average transmitted power can be met with $P_{\text{av}}\leq12$KW when $\theta_{\Delta}={1.39}^\circ$.
Additionally, Fig.\ref{fig:TconofCOSMOS} shows that the $T_{\text{CPI}}$ is shortest in the third visible pass due to its fast target rotation rate. 
Correspondingly, the required average transmitted power in the third visible pass is also the highest. This result can be attributed to two primary factors. First, due to the shorter $T_{\text{CPI}}$ in the third visible pass, a higher $P_{\text{av}}$ is needed to achieve sufficient PSNR, as indicated by (\ref{equ:TconwithE}).
Second, as shown by (\ref{equ:PSNRfucntion}) and (\ref{equ:SRLwithTconandE}), $\rho_{\text{PSNR}\max}$ is inversely proportional to $R^4$, given that $\delta{a}_{\text{SRL}}$ and $\theta_{\Delta}$ are fixed.
Consequently, the larger  $R$ in the third visible pass necessitates a higher  $P_{\text{av}}$. 
As illustrated in Fig.\ref{fig:TconandPavofCOSMOS}, the maximum difference in the required $P_{\text{av}}$ compared to other visible passes can reach nearly an order of magnitude.

Fig.\ref{fig:LEOPav} depicts the tradeoff between $P_{\text{av}}$ and $T_{\text{CPI}}$ at different observing times during the first pass of COSMOS 2494.  
It's evident that the required $P_{\text{av}}$  decreases significantly as $T_{\text{CPI}}$ increases, dropping to as low as a few hundred watts at zenith time.
This reduction is partly due to the increase in $\rho_{\text{PSNR}\max}$ with $T_{\text{CPI}}$. 
More importantly, the LEO satellite's higher rotational rate relative to the radar LOS allows for a larger $\theta_{\Delta}$ to be achieved within a short $T_{\text{CPI}}$.
Therefore, using a longer $T_{\text{CPI}}$ to reduce average transmitted power proves to be a more efficient imaging strategy for LEO satellites. 
Conversely,  increasing $P_{\text{av}}$ can effectively shorten the required $T_{\text{CPI}}$.
For instance, when $P_{\text{av}}$ is increased to 10KW, the required $T_{\text{CPI}}$ drops to below 10s, and even less near zenith time.
However, it is important to note that during the rise and set periods, visibility constraints make it impossible to accumulate sufficient $\theta_{\Delta}$. As a result, a higher $P_{\text{av}}$ is required at these moments, as indicated by the blue circles in Fig.\ref{fig:figPavtotalLEO}.

\vspace{0.1cm}
\noindent {\bf{MEO satellite: NAVSTAR 81}}
\vspace{0.1cm}

\begin{figure}[!t] %t:top，顶部；b：bottom，底部；h：here，当前位置;p:page,浮动页。!:忽视大部分对浮动体位置的内建限制，尽可能地满足我们的位置要求
	\centering
	\subfloat[]{\includegraphics[width=2in]{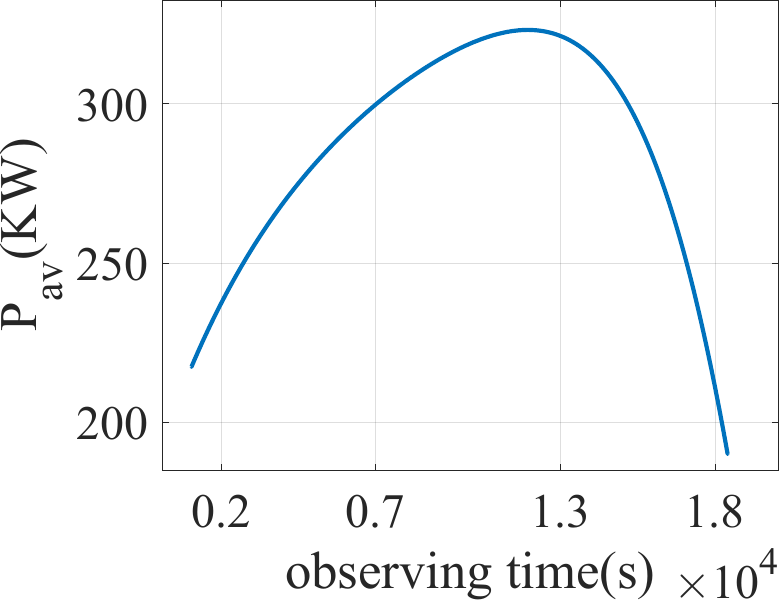}%
		\label{fig:pavofMEO}}
	\hfil
	\subfloat[]{\includegraphics[width=2in]{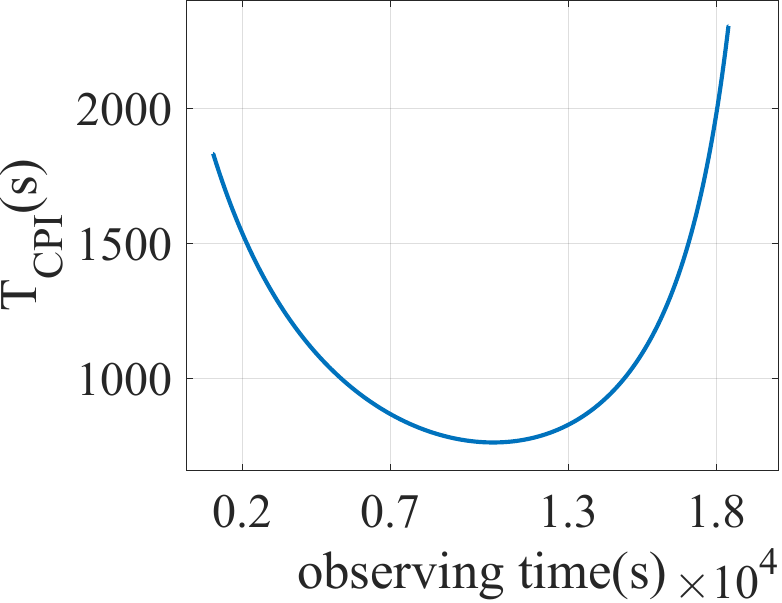}%
		\label{fig:TconofMEO}}
	\caption{the minimum requirement for $P_{\text{av}}$ and $T_{\text{CPI}}$ on imaging NAVSTAR 81. (a) the required $P_{\text{av}}$. (b) the required $T_{\text{CPI}}$.}
	\label{fig:TconandPavofMEO}
	
	\vspace{0.2cm} % 在两部分之间增加垂直间距

	\subfloat[]{\includegraphics[width=2.5in]{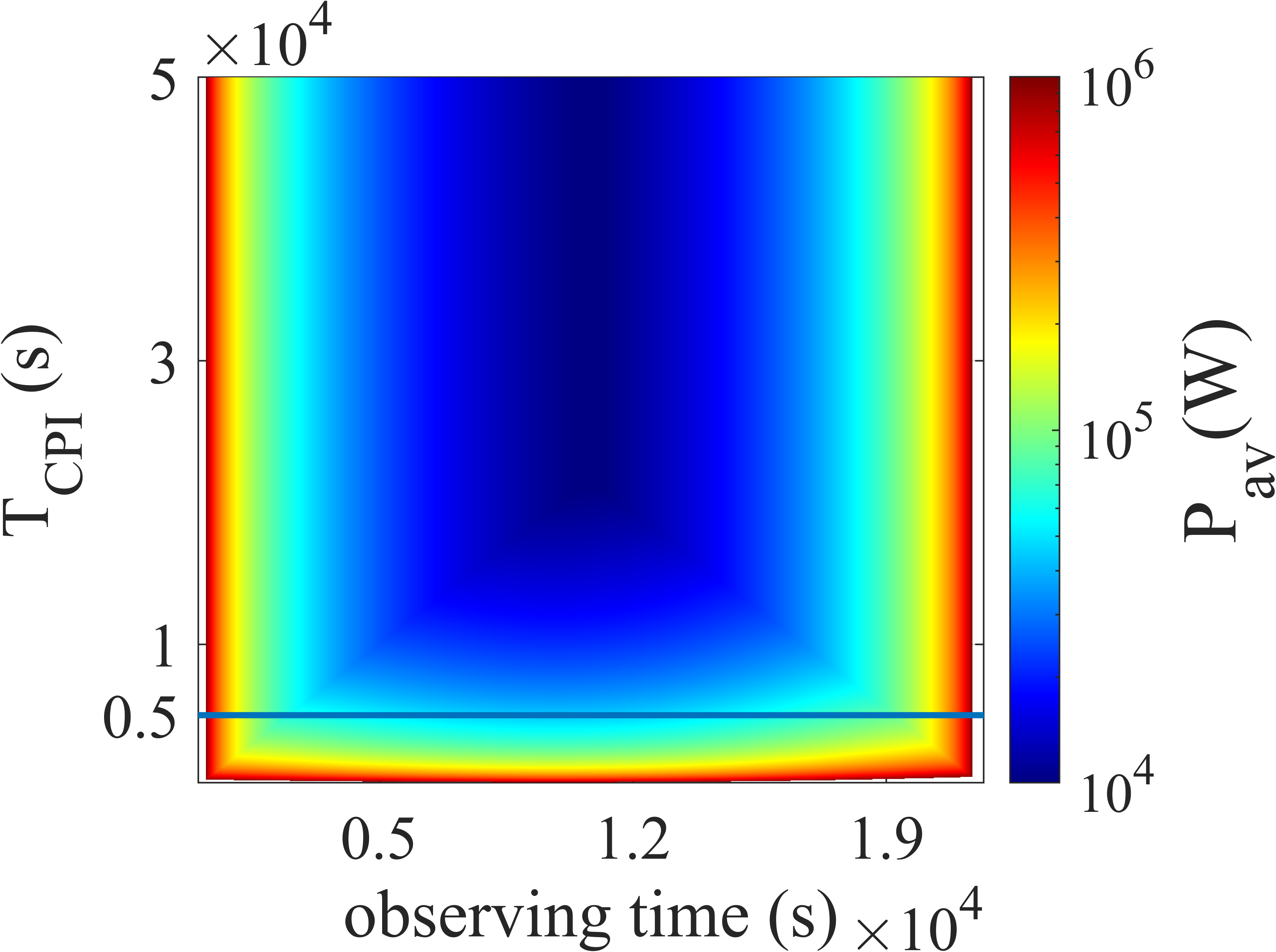}		\label{fig:figPavtotalMEO}}
	\hfil
	\subfloat[]{\includegraphics[width=2.5in]{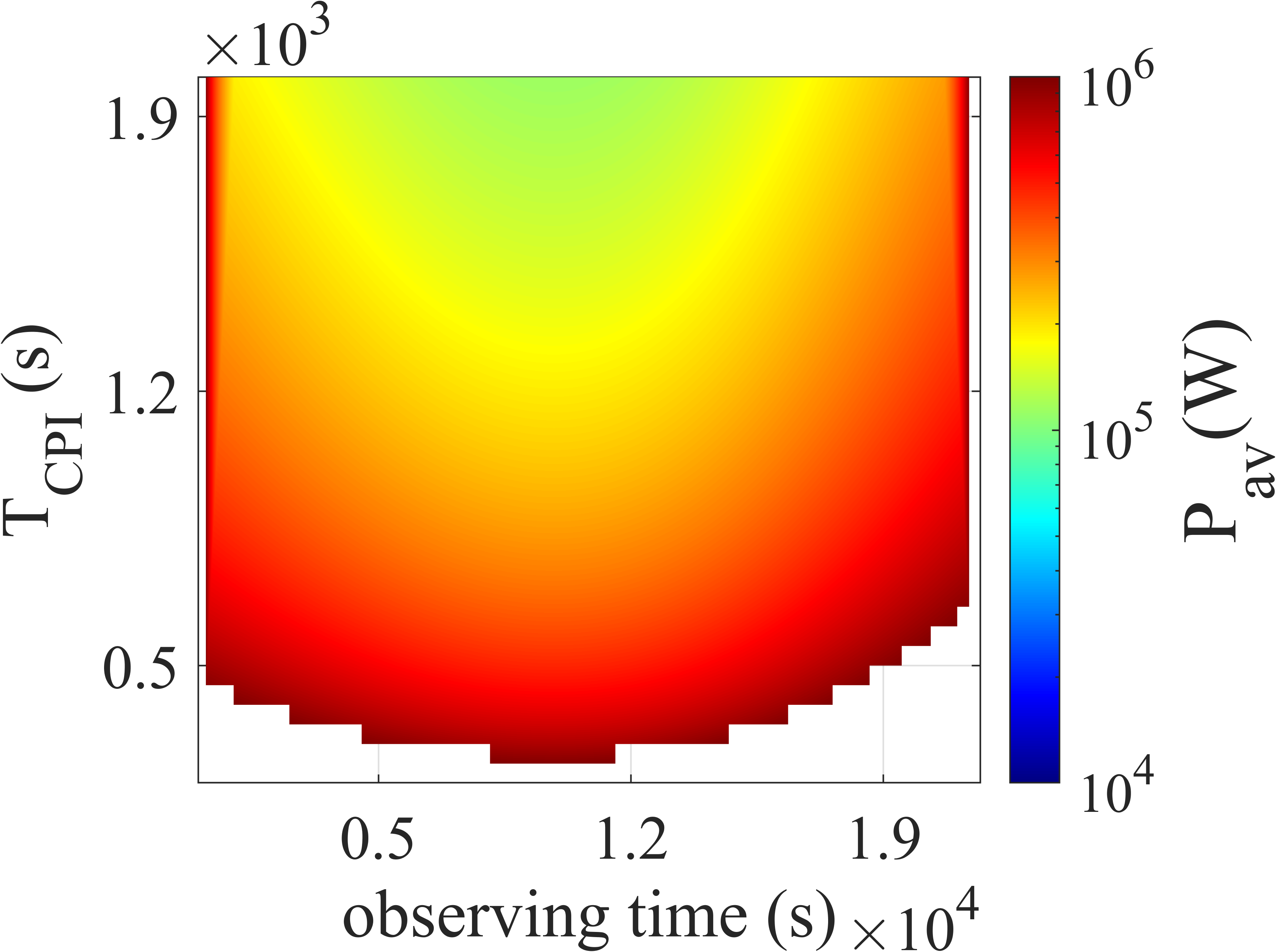}%
		\label{fig:figPavlocalMEO}}
	\caption{the tradeoff between $P_{\text{av}}$ and $T_{\text{CPI}}$ with constant $\delta{a}_{\text{SRL}}$  (pass 1). (a) the total $P_{\text{av}}$. (b) the partial $P_{\text{av}}$.}
	\label{fig:MEOPav}
\end{figure}

The NAVSTAR 81 satellite is selected as the representative MEO satellite in this simulation. Its orbital elements are listed in Table \ref{tabel:radarparameter}. There is only one visible pass within 12 hours starting from 10:00:00 on July 5, 2024, lasting until 16:01:34 on the same day.

Fig.\ref{fig:TconandPavofMEO} illustrates the minimum $P_{\text{av}}$ and $T_{\text{CPI}}$ required for imaging NAVSTAR 81. 
As shown in Fig.\ref{fig:pavofMEO}, the required $P_{\text{av}}$ ranges from 190KW to 320KW when $\theta_{\Delta}=1.39^{\circ}$, which imposes a significant burden on wideband imaging radar systems \cite{macdonald2024overview}. Additionally, the $T_{\text{CPI}}$, depicted in Fig.\ref{fig:TconofMEO}, required to achieve the same $\theta_{\Delta}$ for an MEO satellite is significantly higher than for a LEO satellite, reaching up to 2300s during the rise and set periods.

Fig.\ref{fig:MEOPav} shows the tradeoff between $P_{\text{av}}$ and $T_{\text{CPI}}$ when imaging the NAVSTAR 81 during the visible pass. It's evident that employing a long $T_{\text{CPI}}$ to reduce $P_{\text{av}}$ remains an effective imaging strategy for MEO satellites. 
When $T_{\text{CPI}}$ increases to 5000s, as shown by the line in Fig.\ref{fig:figPavtotalMEO}, the required $P_{\text{av}}$ at most observation times stays within 20KW. While reducing $T_{\text{CPI}}$ leads to a dramatic increase in $P_{\text{av}}$, once $T_{\text{CPI}}$ drops below 2000s, $P_{\text{av}}$ rapidly rises from 100KW to 1MW. If $T_{\text{CPI}}$ falls below 300s, imaging the target becomes impossible, even with $P_{\text{av}}$ as high as 1MW.

\vspace{0.1cm}
\noindent {\bf{IGSO satellite: BeiDou 9}}
\vspace{0.1cm}

This simulation selects the BeiDou 9 satellite as a representative example of IGSO satellites. Its orbital elements are provided in Table \ref{tabel:radarparameter}. There is only one visible arc within 24-hour starting from 15:36:12 on July 4, and lasting until 15:28:00 on July 5.

\begin{figure}[!t] %t:top，顶部；b：bottom，底部；h：here，当前位置;p:page,浮动页。!:忽视大部分对浮动体位置的内建限制，尽可能地满足我们的位置要求
	\centering
	\subfloat[]{\includegraphics[width=2in]{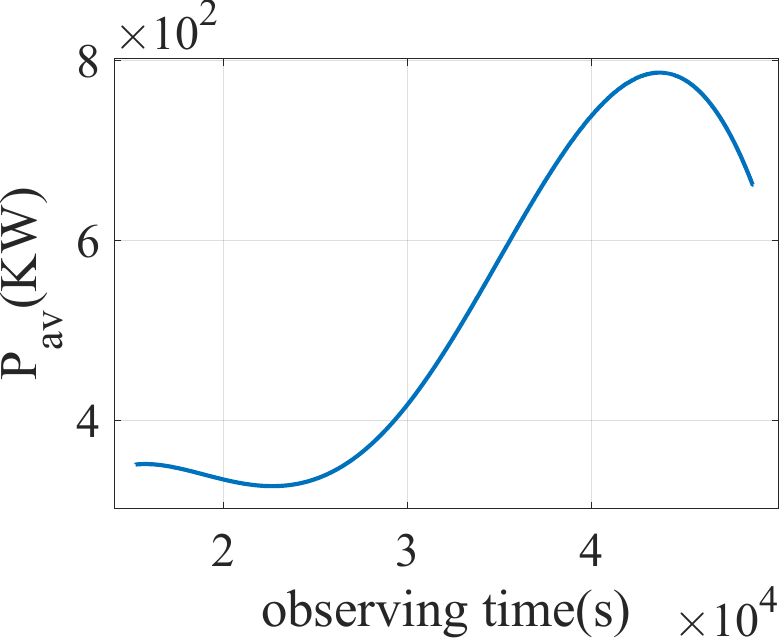}%
		\label{fig:pavofGEO}}
	\hfil
	\subfloat[]{\includegraphics[width=2in]{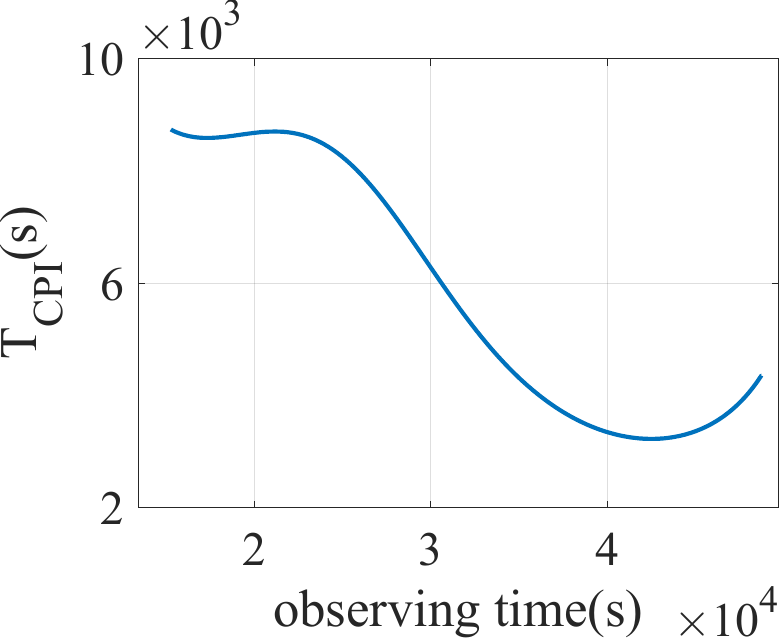}%
		\label{fig:TconofGEO}}
	\caption{the minimum $P_{\text{av}}$ and $T_{\text{CPI}}$ on imaging BeiDou 9. (a) the required $P_{\text{av}}$. (b) the required $T_{\text{CPI}}$.}
	\label{fig:TconandPavofGEO}
     \vspace{0.2cm}
	\centering
	\subfloat[]{\includegraphics[width=2.5in]{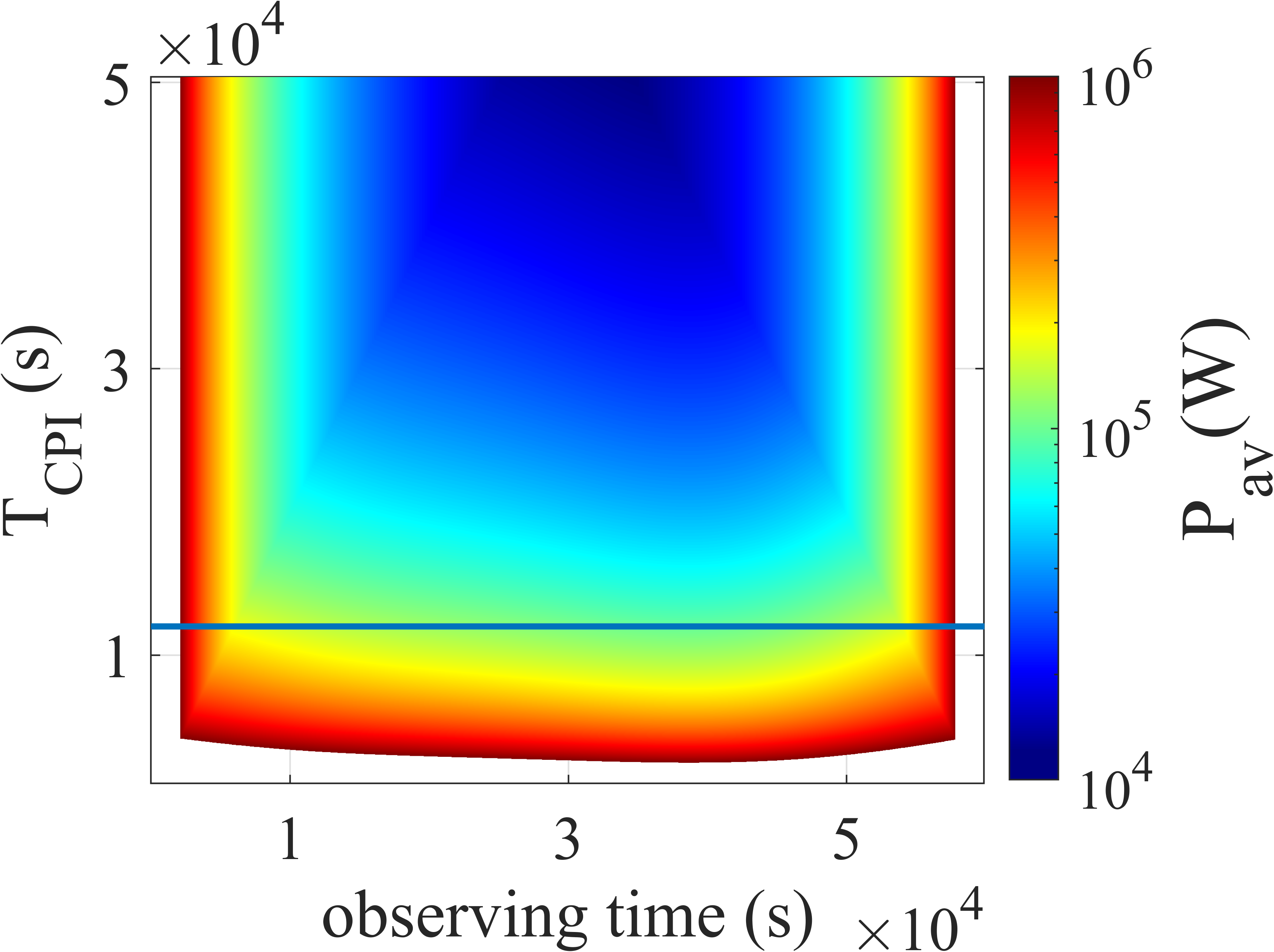}%
		\label{fig:figPavtotalGEO}}
	\hfil
	\subfloat[]{\includegraphics[width=2.5in]{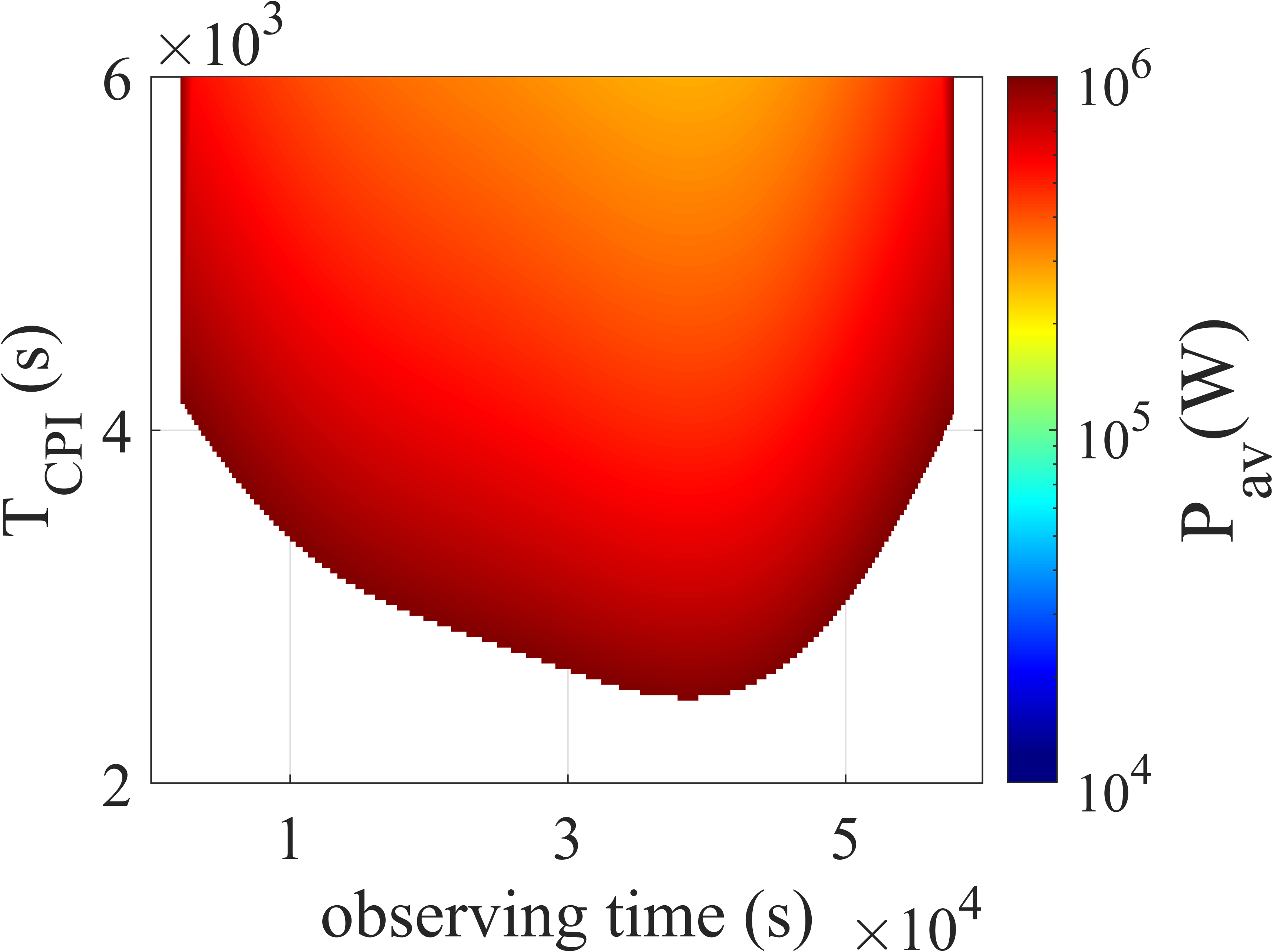}%
		\label{fig:figPavlocalGEO}}
	\caption{the tradeoff between $P_{\text{av}}$ and $T_{\text{CPI}}$ under specified super-resolution limit. (a) the total $P_{\text{av}}$. (b) the partial $P_{\text{av}}$.}
	\label{fig:GEOPav}
\end{figure}

Fig.\ref{fig:TconandPavofGEO} shows the requirements for $P_{\text{av}}$ and $T_{\text{CPI}}$ at each observing time for BeiDou 9.
As illustrated in Fig.\ref{fig:pavofGEO}, the required $P_{\text{av}}$ during this visible pass ranges from 320KW to 790KW when $\theta_{\Delta}=1.39^{\circ}$, exceeding the average transmitted power capabilities of most wideband imaging radars.  Fig.\ref{fig:TconofGEO} shows that the required $T_{\text{CPI}}$ ranges from 3000s to 9000s.

Fig.\ref{fig:GEOPav} presents the tradeoff between $P_{\text{av}}$ and $T_{\text{CPI}}$ when imaging BeiDou 9 during the visible pass. It's apparent that when $T_{\text{CPI}}$ reaches approximately 12000s, as shown by the line in Fig.\ref{fig:figPavtotalGEO}, the required $P_{\text{av}}$ during most imaging time can be reduced to 200KW, which within the detection range of some imaging radars\cite{macdonald2024overview}. Additionally, when $P_{\text{av}}$ reaches 1MW, the $T_{\text{CPI}}$ required in most imaging time can be reduced to less than 4000s.

Furthermore, Fig.\ref{fig:TconofGEO} shows significant differences in the $T_{\text{CPI}}$ required by the target, indicating the presence of more efficient imaging arcs. As shown in Fig.\ref{fig:figPavlocalGEO}, when imaging occurs at approximately 40000s relative to the start of this pass, the required $P_{\text{av}}$ is about 60$\%$ of that during rise and set periods. Although the relative change in resource demand is smaller than that for an LEO target, the extremely high resource demand for imaging GEO targets makes this difference considerable.

It is important to note that the IGSO target selected in this study is a high-inclination IGSO satellite, which can meet imaging requirements through long-term accumulation. There is a compromise between $P_{\text{av}}$ and $T_{\text{CPI}}$. For a geostationary orbit with small inclination and eccentricity, where the target does not rotate relative to the radar LOS, ISAR imaging becomes infeasible.

\section{Conclusion}\label{section:conclusion}

This paper investigates the performance limits of the ISAR super-resolution imaging algorithm for space targets. It derives mathematical expressions for the upper and lower bounds of the computational resolution limit and establishes the relationships between the algorithm's performance, the traditional Rayleigh limit, the number of scatterers, and the PSNR. Based on these findings, the minimum resource requirements, the tradeoffs between these factors, as well as the constraints imposed by azimuth cumulative rotation angle and radar average transmitted power on imaging performance, are analyzed.

Simulation results demonstrate that, for space target imaging, the Rayleigh limit of cross-range resolution in ISAR images typically lies between the upper and lower bounds of the super-resolution imaging algorithm. The super-resolution upper bound falls below the Rayleigh limit only when the number of scatterers is small, and their PSNR is high. Additionally, a tradeoff exists between the azimuth cumulative rotation angle and radar transmitted energy, indicating that imaging efficiency can be significantly improved by increasing the azimuth cumulative rotation angle while reducing the radar's duty cycle. Finally, for a given resolution, the tradeoff between imaging CPI length and average transmitted power, and its impact on imaging efficiency, are highly dependent on the specific imaging scenario.

\section*{Acknowledgments}
Thanks to WenShuo Qian and HaiChen Hu for the discussions and technical assistance, and to the anonymous reviewers for their careful reviews and constructive comments.

\appendices
{\section{}
\label{appendix:appendixtheta}

To simplify the derivation, the Earth is approximated as an ideal sphere, and the relative motion between the radar and the target is modeled as a two-body system. The orbits of space targets are assumed to be perfectly circular, and the targets are considered to maintain a stable three-axis attitude relative to the Earth throughout their orbital motion.
Let $\theta_L$, $\phi$, and $H$ represent the longitude, latitude, and altitude of the target,  respectively, with $R_e$ denoting the Earth's radius. The distance from the target to the Earth's center, $R_c$, is given by $R_c = R_e + H$. 

Assume $t_1$ is the start of the observation period and $t_2 = t_1 + T_{\text{CPI}}$ is the end of the observation period. Then the true anomaly $\theta(t_2)$ and the Greenwich sidereal time $\theta_G(t_2)$ can be approximated as follows:

\begin{equation}
	\begin{aligned}
		& \theta \left( {{t}_{2}} \right)\approx \theta \left( {{t}_{1}} \right)+{{n}_{s}}{{T}_{\text{CPI}}} \\ 
		& {{\theta }_{G}}\left( {{t}_{2}} \right)\approx {{\theta }_{G}}\left( {{t}_{1}} \right)+{{n}_{G}}{{T}_{\text{CPI}}} \\ 
	\end{aligned}
	\label{eq:35}
\end{equation}
where $n_s=\frac{d\theta }{dt}$ represents the angular velocity of the target's orbital motion, and $n_G=\frac{d{{\theta }_{G}}}{dt}$ represents the angular velocity of the Earth's rotation.

Let $\theta$, $a$, $e$, $i$, $\Omega$, and $\omega$ represent the true anomaly, semi-major axis, eccentricity, inclination, longitude of the ascending node, and argument of periapsis, respectively. Using the law of cosines, the distance $R_{\text{CA}}$ from the radar to the target centroid can be formulated as follows:

\begin{equation}
	{{R}_{\text{CA}}}=\sqrt{{{a}^{2}}+R_{c}^{2}-2a{{R}_{c}}\cos \eta }
	\label{eq:36}
\end{equation}
where
\begin{equation*}
	\begin{aligned}
			\cos \eta 
			& =\left[ \cos \Omega \cos \left( \omega +\theta  \right)-\sin \Omega \cos i\sin \left( \omega +\theta  \right) \right]\cos \left( {{\theta }_{G}}+{{\theta }_{L}} \right)\cos \varphi  \\ 
			& +\left[ \sin \Omega \cos \left( \omega +\theta  \right)+\cos \Omega \cos i\sin \left( \omega +\theta  \right) \right]\sin \left( {{\theta }_{G}}+{{\theta }_{L}} \right)\cos \varphi  \\ 
			& +\sin i\sin \left( \omega +\theta  \right)\sin \varphi  \\ 
		\end{aligned}
	\nonumber
\end{equation*}

In the Earth-centered inertial (ECI) coordinate system, the position vectors of the space target, $\vec{A}_{\text{ECI}}$, and the radar station, $\vec{C}_{\text{ECI}}$, can be expressed as follows:

\begin{equation}
	{{\vec{A}}_{\text{ECI}}}={{R}_{z}}\left( \Omega  \right){{R}_{x}}\left( i \right){{R}_{z}}\left( \omega  \right){{\left[ \begin{matrix}
				a\cos \theta  & a\sin \theta  & 0  \\
			\end{matrix} \right]}^{T}}	
	\label{eq:37}
\end{equation}
\begin{equation}
	{{\vec{C}}_{\text{ECI}}}={{R}_{z}}\left( {{\theta }_{G}}+{{\theta }_{L}} \right){{\left[ \begin{matrix}
				{{R}_{c}}\cos \varphi  & 0 & {{R}_{c}}\sin \varphi   \\
			\end{matrix} \right]}^{T}}	
	\label{eq:38}
\end{equation}
where $R_x\left(\cdot \right)$, $R_y\left(\cdot \right)$ and $R_z\left(\cdot \right)$ denote the rotation matrices around the x-axis, y-axis, and z-axis, respectively. A positive Euler angle indicates a counterclockwise rotation along the radar LOS, while a negative Euler angle indicates a clockwise rotation along the radar LOS.

Therefore. in the ECI coordinate system, the radar LOS vector $\vec{E}_{\text{CA-ECI}}\left(t_1\right)$ and the radar LOS velocity vector $\vec{V}_{\text{CA-ECI}}\left(t_1\right)$ can be expressed as:

\begin{equation}
		\begin{aligned}
			& {{{\vec{E}}}_{CA-ECI}}\left( {{t}_{1}} \right)={{{\vec{A}}}_{ECI}}\left( {{t}_{1}} \right)-{{{\vec{C}}}_{ECI}}\left( {{t}_{1}} \right) \\ 
			& =\left[ \begin{matrix}
				{{a}_{11}}\cos \left( \omega +\theta \left( {{t}_{1}} \right) \right)-{{a}_{12}}\sin \left( \omega +\theta \left( {{t}_{1}} \right) \right)-{{a}_{13}}\cos \left( {{\theta }_{G}}\left( {{t}_{1}} \right)+{{\theta }_{L}} \right)  \\
				{{a}_{21}}\cos \left( \omega +\theta \left( {{t}_{1}} \right) \right)+{{a}_{22}}\sin \left( \omega +\theta \left( {{t}_{1}} \right) \right)-{{a}_{23}}\sin \left( {{\theta }_{G}}\left( {{t}_{1}} \right)+{{\theta }_{L}} \right)  \\
				{{a}_{31}}\sin \left( \omega +\theta \left( {{t}_{1}} \right) \right)-{{a}_{32}}  \\
			\end{matrix} \right] \\ 
		\end{aligned}	
	\label{eq:39}
\end{equation}
\begin{equation}
	\begin{aligned}
		{{{\vec{V}}}_{CA-ECI}}\left( {{t}_{1}} \right)
     =\frac{d{{{\vec{E}}}_{CA-ECI}}}{dt}  
	 ={{\mathbf{h}}_{1}}{{n}_{s}}+{{\mathbf{h}}_{2}}{{n}_{G}}  
	\end{aligned}
	\label{eq:40}
\end{equation}
where
\begin{align*}
	a_{11} &= a \cos \left(\Omega\right), 
	a_{12} = a \sin \left(\Omega\right) \cos \left(i\right),
	a_{13} = R_c \cos \left(\phi\right)  \\
	a_{21} &= a \sin \left(\Omega\right), a_{22} = a \cos \left(\Omega\right) \cos \left(i\right), a_{23} = a_{13} 
\end{align*}

\begin{equation}
	\begin{aligned}
		\mathbf{h}_{1} = \left[ \begin{matrix}
			a_{11} \cos \left( \omega + \theta \left( t_1 \right) + \frac{\pi}{2} \right) - a_{12} \sin \left( \omega + \theta \left( t_1 \right) + \frac{\pi}{2} \right) \\
			a_{21} \cos \left( \omega + \theta \left( t_1 \right) + \frac{\pi}{2} \right) + a_{22} \sin \left( \omega + \theta \left( t_1 \right) + \frac{\pi}{2} \right) \\
			a_{31} \sin \left( \omega + \theta \left( t_1 \right) + \frac{\pi}{2} \right)
		\end{matrix} \right]  \nonumber 
	\end{aligned} 
\end{equation}

\begin{equation}
	\begin{aligned}
		\mathbf{h}_{2} = \left[ \begin{matrix}
			-a_{13} \cos \left( \theta_G \left( t_1 \right) + \theta_L + \frac{\pi}{2} \right) \\
			-a_{23} \sin \left( \theta_G \left( t_1 \right) + \theta_L + \frac{\pi}{2} \right) \\
			0
		\end{matrix} \right] \nonumber
	\end{aligned}
\end{equation}

By combining equation (\ref{eq:35}) with trigonometric identities and the Taylor expansion, the following simplified relationships can be obtained:

\begin{equation}
	\begin{aligned}
		\sin & \left( \omega +\theta \left( {{t}_{2}} \right) \right) =\sin \left( \omega +\theta \left( {{t}_{1}} \right)+{{n}_{s}}{{T}_{CPI}} \right) \\
		& \approx \sin \left( \omega +\theta \left( {{t}_{1}} \right) \right)+{{n}_{s}}{{T}_{CPI}}\cos \left( \omega +\theta \left( {{t}_{1}} \right) \right)
	\end{aligned}
	\label{eq:41}
\end{equation}                                      \begin{equation}
	\begin{aligned}
		\cos & \left( \omega +\theta \left( {{t}_{2}} \right) \right)=\cos \left( \omega +\theta \left( {{t}_{1}} \right)+{{n}_{s}}{{T}_{CPI}} \right) \\
		& \approx \cos \left( \omega +\theta \left( {{t}_{1}} \right) \right)-{{n}_{s}}{{T}_{CPI}}\sin \left( \omega +\theta \left( {{t}_{1}} \right) \right)
	\end{aligned}
	\label{eq:42}
\end{equation} 
\begin{equation}
	\begin{aligned}
		\sin & \left( {{\theta }_{G}}\left( {{t}_{2}} \right)+{{\theta }_{L}} \right)=\sin \left( {{\theta }_{G}}\left( {{t}_{1}} \right)+{{n}_{G}}{{T}_{CPI}}+{{\theta }_{L}} \right)\\ 
		& \approx \sin \left( {{\theta }_{G}}\left( {{t}_{1}} \right)+{{\theta }_{L}} \right)+{{n}_{G}}{{T}_{CPI}}\cos \left( {{\theta }_{G}}\left( {{t}_{1}} \right)+{{\theta }_{L}} \right)
	\end{aligned}
	\label{eq:43}
\end{equation} 
\begin{equation}
	\begin{aligned}
		\cos & \left( {{\theta }_{G}}\left( {{t}_{2}} \right)+{{\theta }_{L}} \right)=\cos \left( {{\theta }_{G}}\left( {{t}_{1}} \right)+{{n}_{G}}{{T}_{CPI}}+{{\theta }_{L}} \right) \\
		& \approx \cos \left( {{\theta }_{G}}\left( {{t}_{1}} \right)+{{\theta }_{L}} \right)-{{n}_{G}}{{T}_{CPI}}\sin \left( {{\theta }_{G}}\left( {{t}_{1}} \right)+{{\theta }_{L}} \right)
	\end{aligned}
	\label{eq:44}
\end{equation}                                                                   

Substituting (\ref{eq:41})$\sim$(\ref{eq:44}) into (\ref{eq:39}) fields the radar LOS vector $\vec{E}_{\text{CA-ECI}}\left(t_2\right)$ at $t_2$ moment:

\begin{equation}
	\begin{aligned}	
		\vec{E}_{\text{CA-ECI}}\left(t_2\right) = \vec{E}_{\text{CA-ECI}}\left(t_1\right) + T_{\text{CPI}}\cdot \left[ \begin{matrix}
			{{b}_{1}}  \\
			{{b}_{2}}  \\
			{{b}_{3}}  \\
		\end{matrix} \right]
	\end{aligned}
	\label{eq:45}
\end{equation}
where the series of $b$ is defined as follows:
\begin{equation}
    \begin{aligned}	
			\left[ \begin{matrix}
				{{b}_{1}}  \\
				{{b}_{2}}  \\
				{{b}_{3}}  \\
			\end{matrix} \right]=\left[ \begin{matrix}
				-{{a}_{11}}{{n}_{s}}\sin \left( \omega +\theta \left( {{t}_{1}} \right) \right)-{{a}_{12}}{{n}_{s}}\cos \left( \omega +\theta \left( {{t}_{1}} \right) \right)+{{a}_{13}}{{n}_{G}}\sin \left( {{\theta }_{G}}\left( {{t}_{1}} \right)+{{\theta }_{L}} \right)  \\
				-{{a}_{21}}{{n}_{s}}\sin \left( \omega +\theta \left( {{t}_{1}} \right) \right)+{{a}_{22}}{{n}_{s}}\cos \left( \omega +\theta \left( {{t}_{1}} \right) \right)-{{a}_{23}}{{n}_{G}}\cos \left( {{\theta }_{G}}\left( {{t}_{1}} \right)+{{\theta }_{L}} \right)  \\
				{{a}_{31}}{{n}_{s}}\cos \left( \omega +\theta \left( {{t}_{1}} \right) \right)  \\
			\end{matrix} \right] \notag
    \end{aligned} 
    \label{equ:defiofb}
\end{equation}

The stellar coordinate system, $T_{\text{SCF}}$, is defined as a coordinate system fixed to the space target, with the target's centroid as the origin. The X-axis points from the origin to the geocenter, the Y-axis lies in the orbital plane and aligns with the direction of the target's velocity, and the Z-axis is normal TO the orbital plane. The coordinate transformation matrix $M_{\text{ECI-SCF}}$, which converts from the ECI coordinate system ($T_{\text{ECI}}$) to the stellar coordinate system ($T_{\text{SCF}}$), is given by:
\begin{small}
\begin{equation}
		\begin{aligned}	
			 & {{\mathbf{M}}_{\text{ECI-SCF}}} ={{R}_{z}}\left( -\omega -\theta  \right){{R}_{x}}\left( -i \right){{R}_{z}}\left( -\Omega  \right) \\ 
			& \quad =\left[ \begin{matrix}
				\begin{aligned}
					& \cos \left( -\Omega  \right)\cos \left( -\omega -\theta  \right) \\ 
					& -\sin \left( -\Omega  \right)\cos \left( -i \right)\sin \left( -\omega -\theta  \right) \\ 
				\end{aligned} & \begin{aligned}
					& -\sin \left( -\Omega  \right)\cos \left( -\omega -\theta  \right) \\ 
					& -\cos \left( -\Omega  \right)\cos \left( -i \right)\sin \left( -\omega -\theta  \right) \\ 
				\end{aligned} & \sin \left( -i \right)\sin \left( -\omega -\theta  \right)  \\
				\begin{aligned}
					& \cos \left( -\Omega  \right)\sin \left( -\omega -\theta  \right) \\ 
					& +\sin \left( -\Omega  \right)\cos \left( -i \right)\cos \left( -\omega -\theta  \right) \\ 
				\end{aligned} & \begin{aligned}
					& -\sin \left( -\Omega  \right)\sin \left( -\omega -\theta  \right) \\ 
					& +\cos \left( -\Omega  \right)\cos \left( -i \right)\cos \left( -\omega -\theta  \right) \\ 
				\end{aligned} & -\sin \left( -i \right)\cos \left( -\omega -\theta  \right)  \\
				\sin \left( -\Omega  \right)\sin \left( -i \right) & \cos \left( -\Omega  \right)\sin \left( -i \right) & \cos \left( -i \right)  \\
			\end{matrix} \right] \\ \notag
    \end{aligned} 
    \label{eq:46}
\end{equation}
\end{small}

Substituting (\ref{eq:42})$\sim$(\ref{eq:45}) into (\ref{eq:46}) yields the coordinate transformation matrix at time $t_2$:

\begin{equation}
	{{\mathbf{M}}_{\text{ECI-SCF}}}\left( {{t}_{2}} \right)={{\mathbf{M}}_{\text{ECI-SCF}}}\left( {{t}_{1}} \right)+{{T}_{\text{CPI}}}\cdot \left[ \begin{matrix}
		{{c}_{11}} & {{c}_{12}} & {{c}_{13}}  \\
		{{c}_{21}} & {{c}_{22}} & {{c}_{23}}  \\
		0 & 0 & 0  \\
	\end{matrix} \right]
	\label{eq:47}
\end{equation}
where the series $c$ is defined as:

\begin{small}
	\begin{align}
		& \begin{bmatrix}
				{{c}_{11}} & {{c}_{12}} & {{c}_{13}}  \\
				{{c}_{21}} & {{c}_{22}} & {{c}_{23}}  \\
				0 & 0 & 0  \\
		\end{bmatrix} ={{n}_{s}} \cdot \notag \\
		& \begin{bmatrix}
				\begin{aligned}
					& \cos \left( -\Omega  \right)\sin \left( -\omega -\theta \left( {{t}_{1}} \right) \right) \\ 
					& +\sin \left( -\Omega  \right)\cos \left( -i \right)\cos \left( -\omega -\theta \left( {{t}_{1}} \right) \right) \\ 
				\end{aligned} & \begin{aligned}
					& -\sin \left( -\Omega  \right)\sin \left( -\omega -\theta \left( {{t}_{1}} \right) \right) \\ 
					& +\cos \left( -\Omega  \right)\cos \left( -i \right)\cos \left( -\omega -\theta \left( {{t}_{1}} \right) \right) \\ 
				\end{aligned} & -\sin \left( -i \right)\cos \left( -\omega -\theta \left( {{t}_{1}} \right) \right)  \\
				\begin{aligned}
					& -\cos \left( -\Omega  \right)\cos \left( -\omega -\theta \left( {{t}_{1}} \right) \right) \\ 
					& +\sin \left( -\Omega  \right)\cos \left( -i \right)\sin \left( -\omega -\theta \left( {{t}_{1}} \right) \right) \\ 
				\end{aligned} & \begin{aligned}
					& \sin \left( -\Omega  \right)\cos \left( -\omega -\theta \left( {{t}_{1}} \right) \right) \\ 
					& +\cos \left( -\Omega  \right)\cos \left( -i \right)\sin \left( -\omega -\theta \left( {{t}_{1}} \right) \right) \\ 
				\end{aligned} & -\sin \left( -i \right)\sin \left( -\omega -\theta \left( {{t}_{1}} \right) \right)  \\
				0 & 0 & 0  \\
			\end{bmatrix} \notag
	\end{align}
\end{small}

According to (\ref{eq:47}), the radar LOS vector in $T_{\text{SCF}}$ at $t_2$ can be formulated as follows:
\begin{equation}
		\begin{aligned}
			& {{{\vec{E}}}_{\text{CA-SCF}}}\left( {{t}_{2}} \right)  ={{\mathbf{M}}_{\text{ECI-SCF}}}\left( {{t}_{2}} \right){{{\vec{E}}}_{\text{CA-SCF}}}\left( {{t}_{2}} \right) \\ 
			&\quad = \left( {{\mathbf{M}}_{\text{ECI-SCF}}}\left( {{t}_{1}} \right)+{{T}_{\text{CPI}}}\cdot \left[ \begin{matrix}
				{{c}_{11}} & {{c}_{12}} & {{c}_{13}}  \\
				{{c}_{21}} & {{c}_{22}} & {{c}_{23}}  \\
				0 & 0 & 0  \\
			\end{matrix} \right] \right)  \quad \cdot \left( {{{\vec{E}}}_{\text{CA-SCF}}}\left( {{t}_{1}} \right)+{{T}_{\text{CPI}}}\cdot \left[ \begin{matrix}
				{{b}_{1}}  \\
				{{b}_{2}}  \\
				{{b}_{3}}  \\
			\end{matrix} \right] \right) \\ 
			&\quad  =T_{\text{CPI}}^{2}{{{\vec{I}}}_{1}}+{{T}_{\text{CPI}}}{{{\vec{I}}}_{2}}+{{{\vec{E}}}_{\text{CA-SCF}}}\left( {{t}_{1}} \right) \\ 
		\end{aligned}	
	\label{eq:48}
\end{equation}
where
\begin{equation}
	{{\vec{I}}_{1}}=\left[ \begin{matrix}
		{{c}_{11}} & {{c}_{12}} & {{c}_{13}}  \\
		{{c}_{21}} & {{c}_{22}} & {{c}_{23}}  \\
		0 & 0 & 0  \\
	\end{matrix} \right]\cdot \left[ \begin{matrix}
		{{b}_{1}}  \\
		{{b}_{2}}  \\
		{{b}_{3}}  \\
	\end{matrix} \right]=\left[ \begin{matrix}
		{{h}_{1}}  \\
		{{h}_{2}}  \\
		0  \\
	\end{matrix} \right] \nonumber	
\end{equation}
\begin{equation}
	\begin{aligned}
		& {{h}_{1}}={{c}_{11}}{{b}_{1}}+{{c}_{12}}{{b}_{2}}+{{c}_{13}}{{b}_{3}} \\ 
		& {{h}_{2}}={{c}_{21}}{{b}_{1}}+{{c}_{22}}{{b}_{2}}+{{c}_{23}}{{b}_{3}} \\ 
	\end{aligned} \nonumber	
\end{equation}
\begin{equation}
		{{\vec{I}}_{2}}={{\mathbf{M}}_{\text{ECI-SCF}}}\left( {{t}_{1}} \right)\cdot \left[ \begin{matrix}
			{{b}_{1}}  \\
			{{b}_{2}}  \\
			{{b}_{3}}  \\
		\end{matrix} \right]+\left[ \begin{matrix}
			{{c}_{11}} & {{c}_{12}} & {{c}_{13}}  \\
			{{c}_{21}} & {{c}_{22}} & {{c}_{23}}  \\
			0 & 0 & 0  \\
		\end{matrix} \right]\cdot {{\vec{E}}_{\text{CA-ECI}}}\left( {{t}_{1}} \right)=\left[ \begin{matrix}
			{{g}_{1}}  \\
			{{g}_{2}}  \\
			{{g}_{3}}  \\
		\end{matrix} \right] \nonumber
\end{equation}

Assume that the radar LOS vector in the time-invariant reference imaging coordinate system $T_{\text{Imag0}}$ at $t_1$ is given by ${{\vec{E}}_{\text{CA-Img0}}}\left( {{t}_{1}} \right)={{\left[ {{x}_{{{t}_{1}}}},{{y}_{{{t}_{1}}}},{{z}_{{{t}_{1}}}} \right]}^{T}}$. After the imaging accumulation time $T_{\text{CPI}}$,  the radar LOS vector at $t_2$ in $T_{\text{Imag0}}$ is given by ${{\vec{E}}_{\text{CA-Img0}}}\left( {{t}_{2}} \right)={{\left[ {{x}_{{{t}_{2}}}},{{y}_{{{t}_{2}}}},{{z}_{{{t}_{2}}}} \right]}^{T}}$.
During this period, the cumulative rotation angle $\theta_{\Delta}$ can be expressed as:

\begin{equation}
		\begin{aligned}
			{{\theta }_{\Delta }} 
			& =\operatorname{acos}\left( \frac{{{x}_{{{t}_{1}}}}{{x}_{{{t}_{2}}}}+{{y}_{{{t}_{1}}}}{{y}_{{{t}_{2}}}}}{\sqrt{x_{{{t}_{1}}}^{2}+y_{{{t}_{1}}}^{2}}\sqrt{x_{{{t}_{2}}}^{2}+y_{{{t}_{2}}}^{2}}} \right) \\ 
			& =\operatorname{acos}\left( \frac{T_{\text{CPI}}^{2}{{d}_{1}}+{{T}_{\text{CPI}}}{{d}_{2}}+{{d}_{3}}}{\sqrt{T_{\text{CPI}}^{4}{{d}_{4}}+T_{\text{CPI}}^{3}{{d}_{5}}+T_{\text{CPI}}^{2}{{d}_{6}}+{{T}_{\text{CPI}}}{{d}_{7}}+{{d}_{8}}}} \right) \\ 
		\end{aligned} 
	\label{eq:49}
\end{equation}
where

\begin{equation}
	\begin{aligned}
		& {{d}_{1}}={{x}_{{{t}_{1}}}}{{h}_{1}}+{{y}_{{{t}_{1}}}}{{h}_{2}} \\ 
		& {{d}_{2}}={{x}_{{{t}_{1}}}}{{g}_{1}}+{{y}_{{{t}_{1}}}}{{g}_{2}} \\ 
		& {{d}_{3}}=x_{{{t}_{1}}}^{2}+y_{{{t}_{1}}}^{2} \\ 
		& {{d}_{4}}=\left( h_{1}^{2}+h_{2}^{2} \right)\left( x_{{{t}_{1}}}^{2}+y_{{{t}_{1}}}^{2} \right) \\ 
		& {{d}_{5}}=\left( 2{{h}_{1}}{{g}_{1}}+2{{h}_{2}}{{g}_{2}} \right)\left( x_{{{t}_{1}}}^{2}+y_{{{t}_{1}}}^{2} \right) \\ 
		& {{d}_{6}}=\left( g_{1}^{2}+g_{2}^{2}+2{{h}_{1}}{{x}_{{{t}_{1}}}}+2{{h}_{2}}{{y}_{{{t}_{1}}}} \right)\left( x_{{{t}_{1}}}^{2}+y_{{{t}_{1}}}^{2} \right) \\ 
		& {{d}_{7}}=\left( 2{{g}_{1}}{{x}_{{{t}_{1}}}}+2{{g}_{2}}{{y}_{{{t}_{1}}}} \right)\left( x_{{{t}_{1}}}^{2}+y_{{{t}_{1}}}^{2} \right) \\ 
		& {{d}_{8}}={{\left( x_{{{t}_{1}}}^{2}+y_{{{t}_{1}}}^{2} \right)}^{2}} \\ 
	\end{aligned}\nonumber
\end{equation}
}

{\section{}
\label{appendix:appendixPSNR}
The tracking radar range equation from \cite{ref54} is given by:
\begin{align}
    R^4_{\max} = \frac{ P_{\text{av}}  G A_e  \sigma_{\text{RCS}} n E_i(n) F^4 }{ \left( 4\pi \right)^2 k T_0 F_n F_{\text{PRF}} \left( S/N \right)_1 L_s}
\end{align}
where $G$ is the antenna gain on transmitter, $A_e$ is the effective area on receiver, $n$ is the pulse number, $E_i(n)$ is the efficiency in adding together $n$ pulses, $F^4$ represents the propagation effects of electromagnetic waves, $k$ is the Boltzmann's constant, $T_0$ is the standard temperature, $F_n$ is the noise figure of the receiver,  $\left( S/N \right)_1$ stands for the required SNR if only one pulse is present, $L_s$ is the radar system losses.
Let 
\begin{align*}
    C_{\text{st}} =  \frac{ G A_e E_i(n) F^4  }{ \left( 4\pi \right)^2 k T_0 F_n L_s  }
\end{align*}
and $T_{\text{CPI}} = n/ F_{\text{PRF}}$, then the equation (\ref{equ:PSNRfucntion}) is obtained.
}

\bibliographystyle{IEEEtran}
\bibliography{References.bib} %ref为.bib文件名
 
\vspace{11pt}

\begin{IEEEbiography}[{\includegraphics[width=1in,height=1.25in,clip,keepaspectratio]{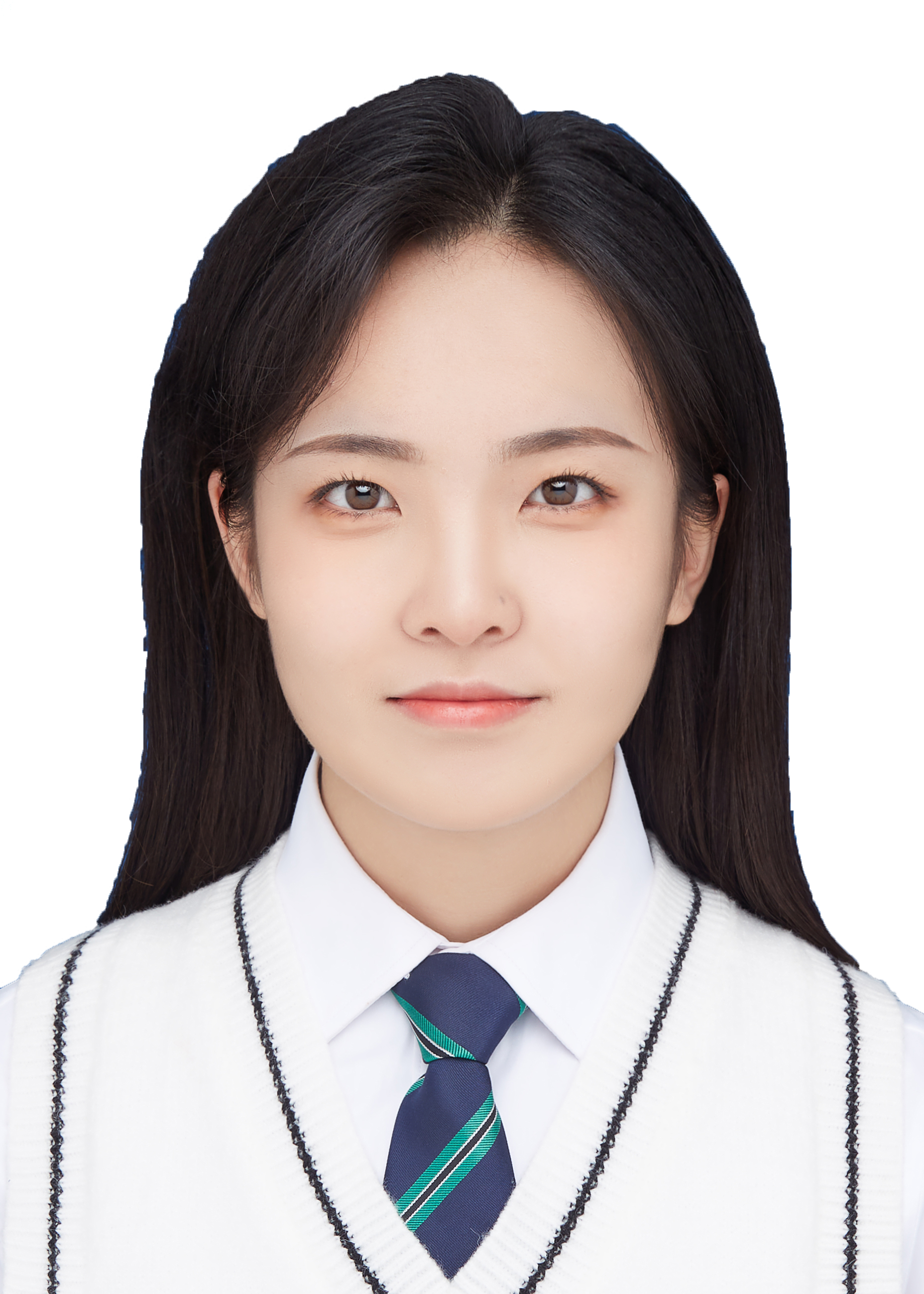}}]{Xiaole He}
was born in Shuozhou, Shanxi, China, in 1999. She received an M.E. degree from the Xidian University, China, in 2021.

Now, she is currently pursuing the Ph.D. degree at Beijing Institute of Technology, Beijing, China. Her research interests include ISAR imaging and super-resolution.
\end{IEEEbiography}
\vspace{-10pt} % 减少间距
\begin{IEEEbiography}[{\includegraphics[width=1in,height=1.25in,clip,keepaspectratio]{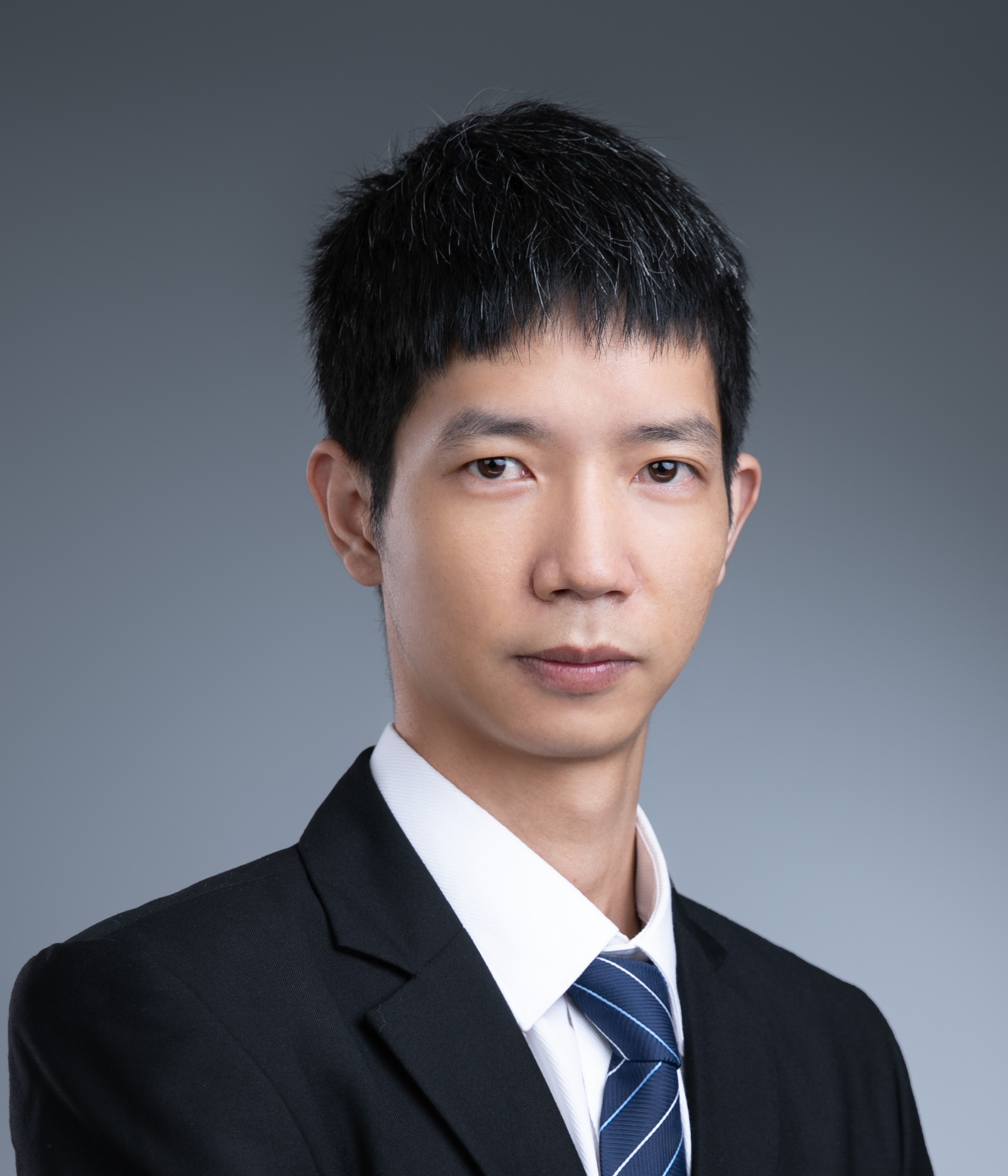}}]{Ping Liu}
	received the B.S. degree in mathematics from Wuhan University in 2016, and the Ph.D. degree in mathematics from the Hong Kong University of Science and Technology in 2021. 
	
	He was a postdoc with the Department of Mathematics, ETH Zürich, from 2021 to 2024. Since 2024, he has been a researcher with the School of Mathematical Sciences, Zhejiang University. His current research interests include super-resolution imaging, array signal processing, topological photonics, and topological phononics. 
\end{IEEEbiography}
\vspace{-10pt} % 减少间距
\begin{IEEEbiography}[{\includegraphics[width=1in,height=1.25in,clip,keepaspectratio]{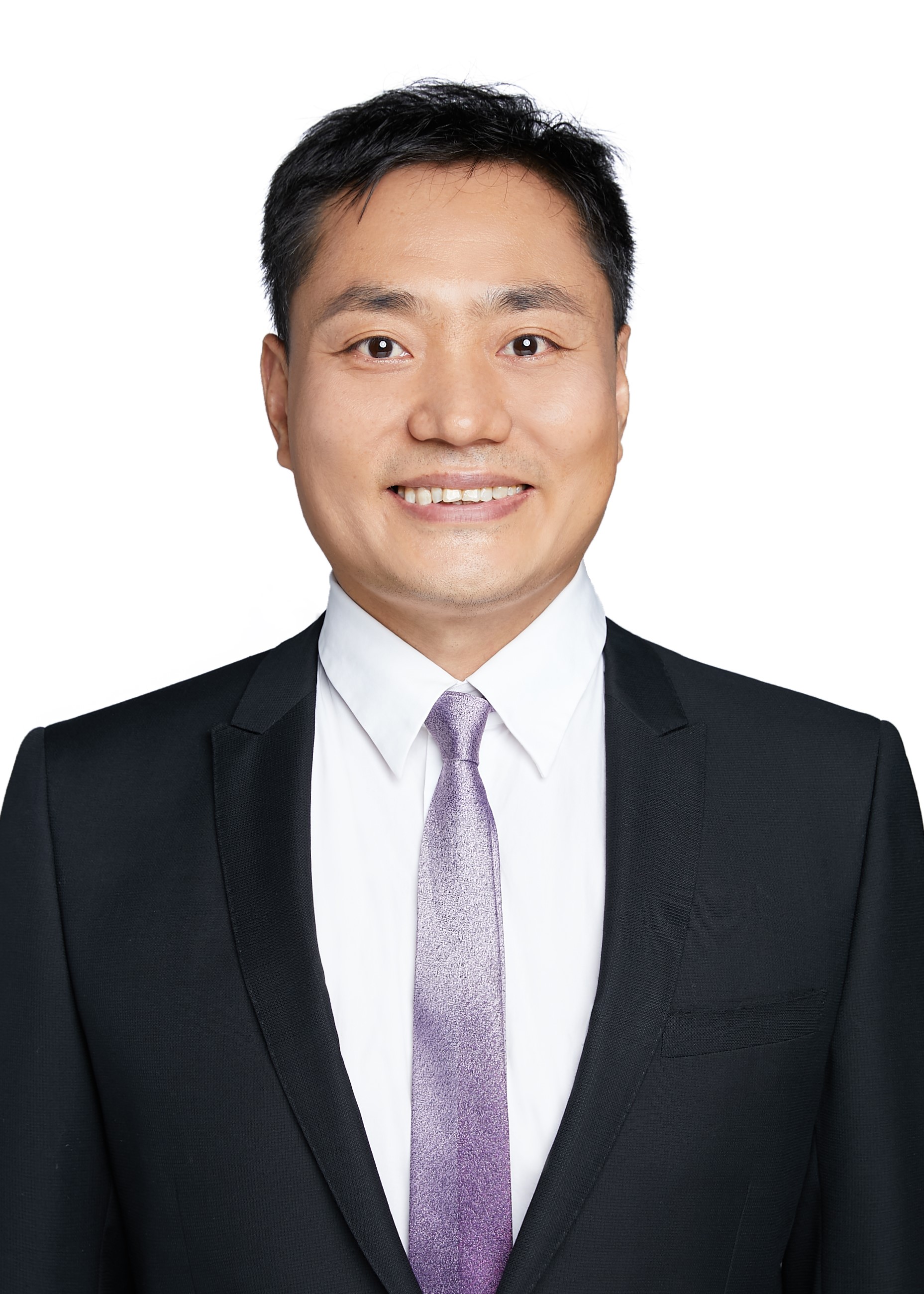}}]{Junling Wang}
   	(Member, IEEE)  received the B.E. and M.S. degrees from China University of Petroleum, Qingdao, China, in 2005 and in 2008, respectively, and the Ph.D. degree from Beijing Institute of Technology (BIT), in 2013. He was an exchange student in the Department of Signal Theory and Communications, Universitat Politecnica de Catalunya in 2010. 
   	
   	Since 2013, he has been working with the School of Information and Electronics, BIT, Beijing, China, where he is currently an Associate Professor. His current research interests include satellite detection and imaging, and radar signal processing.
\end{IEEEbiography}

\vspace{11pt}

\end{document}